\documentclass[twocolumn, 10pt, pra, aps, showpacs, reprint, amsmath, amssymb, superscriptaddress, nofootinbib]{revtex4-1}

\usepackage{times}
\usepackage[pdftex]{graphicx} 

\begin{document}

\title{Suppressing spatio-temporal lasing instabilities with wave-chaotic microcavities}

\author{Stefan Bittner}
\affiliation{Department of Applied Physics, Yale University, New Haven, Connecticut 06520, USA}
\author{Stefano Guazzotti}
\affiliation{Blackett Laboratory, Department of Physics, Imperial College London, London SW7 2AZ, United Kingdom}
\author{Yongquan Zeng}
\author{Xiaonan Hu}
\affiliation{Center for OptoElectronics and Biophotonics, School of Electrical and Electronic Engineering and The Photonics Institute, Nanyang Technological University, 50 Nanyang Avenue, 639798 Singapore}
\author{Hasan Y{\i}lmaz}
\author{Kyungduk Kim}
\affiliation{Department of Applied Physics, Yale University, New Haven, Connecticut 06520, USA}
\author{Sang Soon Oh}
\affiliation{Blackett Laboratory, Department of Physics, Imperial College London, London SW7 2AZ, United Kingdom}
\affiliation{School of Physics and Astronomy, Cardiff University, Cardiff CF24 3AA, United Kingdom}
\author{Qi Jie Wang}
\affiliation{Center for OptoElectronics and Biophotonics, School of Electrical and Electronic Engineering and The Photonics Institute, Nanyang Technological University, 50 Nanyang Avenue, 639798 Singapore}
\author{Ortwin Hess}
\email{o.hess@imperial.ac.uk}
\affiliation{Blackett Laboratory, Department of Physics, Imperial College London, London SW7 2AZ, United Kingdom}
\author{Hui Cao}
\email{hui.cao@yale.edu}
\affiliation{Department of Applied Physics, Yale University, New Haven, Connecticut 06520, USA}

\begin{abstract}
Spatio-temporal instabilities are widespread phenomena resulting from complexity and nonlinearity. In broad-area edge-emitting semiconductor lasers, the nonlinear interactions of multiple spatial modes with the active medium can result in filamentation and spatio-temporal chaos. These instabilities degrade the laser performance and are extremely challenging to control. We demonstrate a powerful approach to suppress spatio-temporal instabilities using wave-chaotic or disordered cavities. The interference of many propagating waves with random phases in such cavities disrupts the formation of self-organized structures like filaments, resulting in stable lasing dynamics. Our method provides a general and robust scheme to prevent the formation and growth of nonlinear instabilities for a large variety of high-power lasers.
\end{abstract}

\maketitle

Systems with complex spatio-temporal dynamics can exhibit instabilities and even chaotic dynamics, as seen for example in weather patterns, turbulent flow, population dynamics \cite{Hassell1991}, or chemical reactions \cite{Epstein1996}. Beyond natural occurrences, spatio-temporal instabilities also appear in sophisticated technological systems such as fusion reactors exhibiting plasma instabilities \cite{Graves2012} or type-II superconductors with complex vortex dynamics \cite{Altshuler2004}. Lasers are another important class of systems exhibiting inherent spatio-temporal instabilities and deterministic chaos due to the nonlinear interaction of the light field with the active medium \cite{Abraham1985,Huyet1995,OthsuboBook2013}. The underlying nonlinearities are particularly pronounced for high power lasers, which have a large transverse area and operate on many spatial (transverse) modes. Nonlinear modal interactions entail spatio-temporal instabilities such as irregular pulsation and filamentation, e.g., in broad-area edge-emitting semiconductor lasers \cite{Fischer1996,Hess1994,Hess1996,Marciante1997,Marciante1998,Scholz2008,Arahata2015}, that degrade the spatial, spectral and temporal properties of the emission. 

Because of wide-spread applications of high power lasers in material processing, large-scale displays, laser surgery and pumping sources, much effort has been invested in suppressing lasing instabilities. Most strategies proposed seek to reduce the level of complexity by reducing the number of lasing modes. For broad-area semiconductor lasers, this can be achieved by external control, e.g., through injection of a coherent signal \cite{Pawletko2000,Takimoto2009a} or delayed optical feedback \cite{Martin-Regalado1996,Mandre2005,Zink2014}, or schemes based on Parity-Time symmetry \cite{Hodaei2014,Liu2017}. Successful with moderate powers, these approaches quickly become less effective when increasing the cavity size in order to harness more power. An external control signal applied via injection or feedback through the cavity boundary has a diminished effect deep inside a large cavity and it thus fails to control the dynamics over the whole cavity. Furthermore, these approaches are typically sensitive to external perturbations and require a careful adjustment of parameters to reach stabilization. 

Our approach aims to eliminate spatio-temporal instabilities in broad-area edge-emitting semiconductor lasers without reducing the number of lasing modes and is thus applicable to high power operation. Instead of suppressing the filaments via external signals, we disrupt the coherent nonlinear processes that lead to their formation by using cavities with complex spatial structure to create many propagating waves with random phases. The complex interference of these waves prevents the formation of self-organized structures such as filaments that are prone to modulational instabilities. We demonstrate the generality and robustness of this approach through experiments and numerical simulations with two different systems, (i) two-dimensional (2D) microcavities featuring chaotic ray dynamics and (ii) one-dimensional (1D) cavities with random fluctuations of the refractive index. The chaotic ray dynamics and the structural disorder are responsible for the creation of multi-wave interference effects, respectively. 

\section*{Conventional broad-area edge-emitting lasers}

\begin{figure*}[tb]
\begin{center}
\includegraphics[width = 13.0 cm]{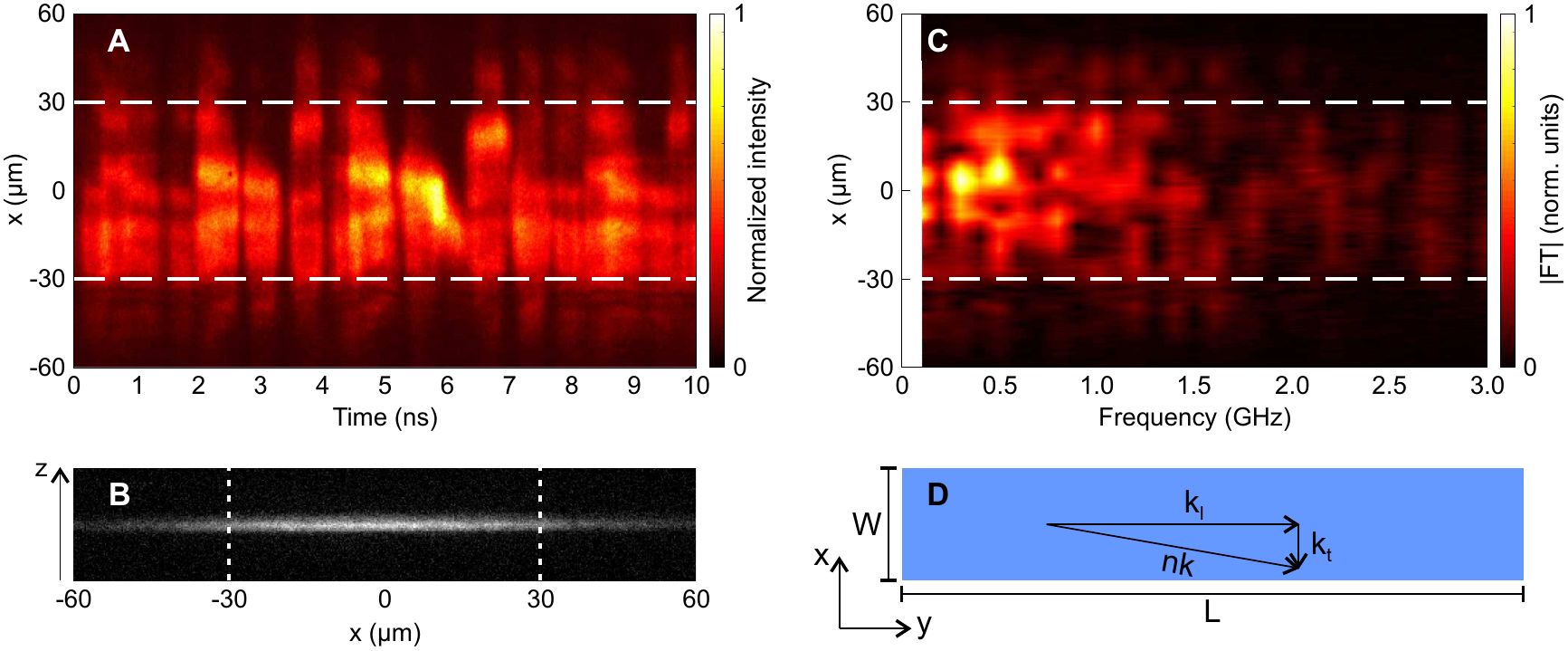}
\end{center}
\caption{Spatio-temporal instabilities of an electrically pumped edge-emitting semiconductor laser. The top metal contact is $60~\mu$m wide and $0.98$~mm long. 
	(\textbf{A})~Spatio-temporal image of the emission intensity $I(x,t)$ at one of the cleaved facets. The stripe laser was injected with an electric current of $400$~mA and $2~\mu$s-long pulses, where the lasing threshold current was $I_{th} = 230$~mA. The image was taken at $0.37~\mu$s after the start of the pump pulse, well beyond any turn-on transient. Part of the emission stems from outside the region of the top contact (marked by white dashed lines) due to lateral spread of the injected current in the GaAs.
	(\textbf{B})~Image of electroluminescence from the end facet for $100$~mA pump current (below threshold) and a pulse length of $t_p = 20~\mu$s. The emission is spatially homogeneous without any visible defects. 
	(\textbf{C})~Temporal Fourier transform of $I(x,t)$ in (A).
  (\textbf{D})~Sketch of a rectangular Fabry-Perot cavity of length $L$ and width $W$ where $L \gg W$. The wave vector can be separated into longitudinal and transverse components $k_l$ and $k_t$, respectively. Since $k_l \gg k_t$, the lasing modes propagate predominantly in the longitudinal ($y$) direction.
}
\label{fig:FPdynamics}
\end{figure*}

We first show the complex spatio-temporal dynamics of GaAs quantum well (QW) lasers in the widely-used stripe geometry. The reflections from two cleaved facets in the longitudinal direction (parallel to the stripe axis) and gain guiding in the transverse direction (perpendicular to the stripe axis) provide optical confinement [see methods \cite{Methods}]. Spatio-temporal traces of the lasing emission intensity at one end facet were measured by a streak camera with picosecond resolution [see methods \cite{Methods}]. As shown in Figure \ref{fig:FPdynamics}A, the lasing emission is spatially concentrated at multiple locations --- so-called filaments --- which sometimes move in the transverse direction \cite{Hess1994,Fischer1996,Hess1996}. Emission patterns measured during the same pulse in Fig.~\ref{fig:FPstreakExmpls} demonstrate that the lasing emission can change suddenly from a nearly uniform distribution to concentration in small regions or filaments. Such diverse emission profiles illustrate that the formation of filaments is an inherent feature of the lasing dynamics and not due to inhomogeneities of the cavity. This is confirmed by an electroluminescence image (Fig.~\ref{fig:FPdynamics}B) taken below lasing threshold displaying a homogeneous intensity distribution across the facet. Furthermore, the lasing emission oscillates rapidly and irregularly in time (Fig.~\ref{fig:FPdynamics}A). The spatially-resolved temporal Fourier transform of the emission intensity $I(x,t)$ (Fig.~\ref{fig:FPdynamics}C) reveals a number of frequency components up to about $1.5$~GHz, which accounts for the irregular oscillations on a nanosecond time scale. 

The filaments are formed through spatio-temporal nonlinear processes including spatial hole burning, carrier-induced index variation and self-focusing \cite{Fischer1996,Hess1994,Hess1996,Marciante1997,Marciante1998,Scholz2008,Arahata2015}. The stripe laser cavity is of Fabry-Perot (FP) type, and the light field propagates predominantly in the longitudinal direction. The wave vector component in the longitudinal direction, $k_l$, is much larger than that in the transverse direction, $k_t$. Consequently, the transverse wavelength $\lambda_t = 2 \pi / k_t$ is typically on the order of a few micrometers, and much longer than the longitudinal wavelength $\lambda_l = 2 \pi / k_l$. A variation of the field intensity in the transverse direction on the scale of $\lambda_t$ can result in filamentation due to carrier-induced index changes: a region of increased intensity depletes the gain, thus raising the refractive index locally and forming a lens. The lens will focus light and further enhance local intensity. This self-focusing process generates a filament with a typical width of several micrometers, comparable to the transverse wavelength. Since the optical gain is less depleted outside the filament, the filament tends to migrate transversely to the neighboring region of higher gain. Meanwhile, additional filaments may form at locations of less carrier depletion. These filaments will interact nonlinearly via the semiconductor quantum well.  Due to dynamic gain and nonlinear interaction, the filaments vary in space and time, leading to complex spatio-temporal dynamics and instabilities \cite{Hess1994}. The resulting degradation and temporal fluctuations of the output profile limit the laser applications. 

\section*{Wave-chaotic microcavity lasers}

Microcavities with chaotic ray dynamics \cite{Tureci2005,Harayama2010,Cao2015} have been studied in the context of wave-dynamical chaos \cite{StoeckmannBuch2000}. The resonant modes of the passive cavities are determined by a linear wave equation and do not exhibit chaos in the sense of an exponential sensitivity to the initial conditions. However, the chaotic ray dynamics manifests in the spatial and spectral properties of the cavity resonances, e.g., the spatial field distributions feature a pseudo-random, speckle-like structure. Such wave-chaotic microcavities have been used to tailor the steady-state lasing properties such as output directionality, lasing threshold and spectrum \cite{Tureci2005,Xiao2010b,Harayama2010,Cao2015,Jiang2017}. Here we investigate the temporal dynamics of such lasers. 

\begin{figure}[tb]
\begin{center}
\includegraphics[width = 8.4 cm]{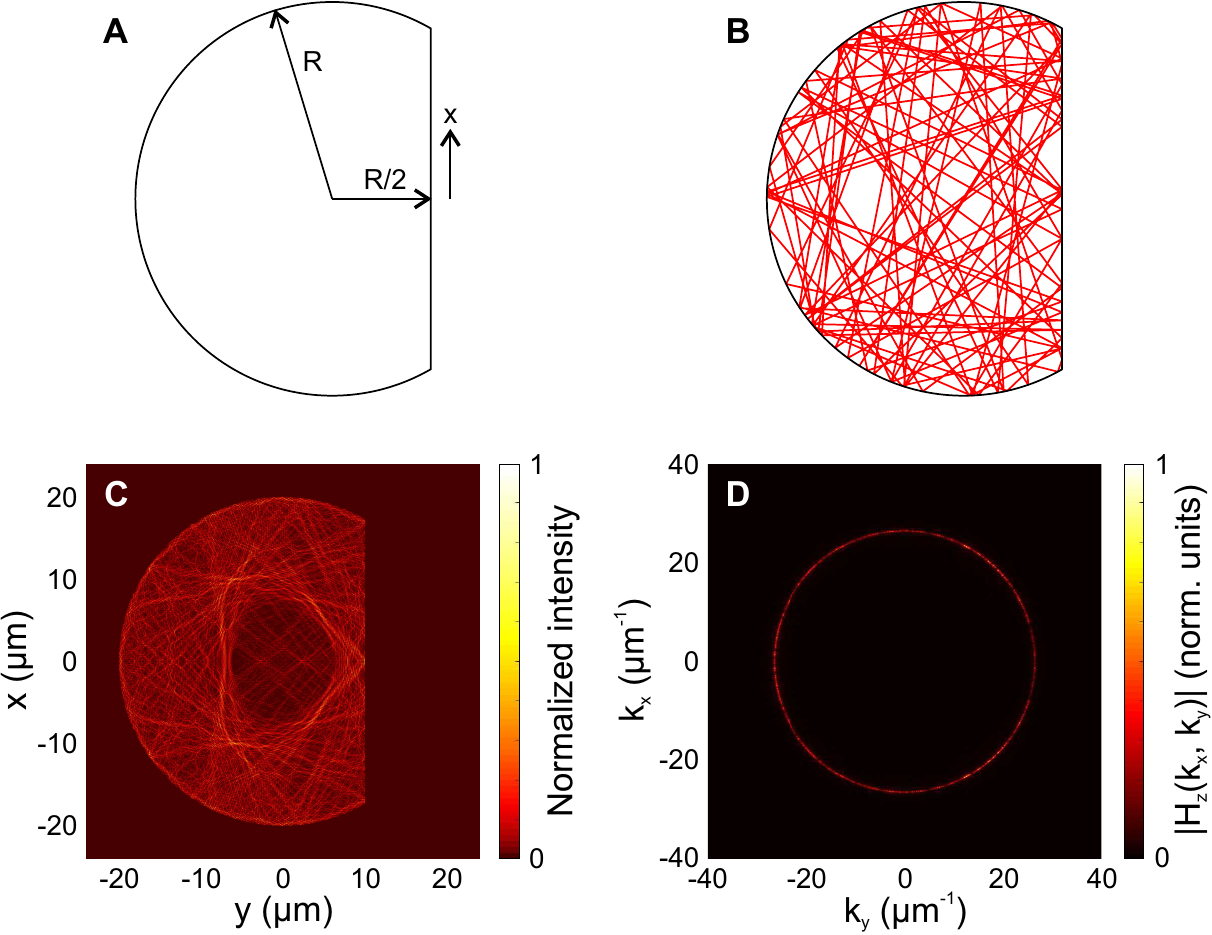}
\end{center}
\caption{D-cavity with chaotic ray dynamics.
	(\textbf{A})~Geometry of the D-cavity. A section is removed from a circle with radius $R$ along a straight cut $R/2$ away from the center. The coordinate along this segment of the boundary is denoted by $x$.
	(\textbf{B})~A typical ray trajectory in a closed D-cavity covers the entire cavity and propagates in all directions. 
	(\textbf{C})~Intensity distribution of a typical high-$Q$ mode ($\lambda = 800.4$~nm, $Q = 3443$) in a dielectric D-cavity with radius $R = 20~\mu$m and refractive index $n = 3.37$.
	(\textbf{D})~The wave-vector distribution of the same mode is isotropic, indicating there is no dominant direction of propagation.}
\label{fig:passiveModes}
\end{figure}

As an example, we consider a D-shaped cavity (Fig.~\ref{fig:passiveModes}A), which has fully chaotic ray dynamics. A single trajectory (Fig.~\ref{fig:passiveModes}B) generally covers the entire cavity and propagates in all possible directions. The classical ray dynamics manifests in the spatial structure of the resonant modes (Fig.~\ref{fig:passiveModes}C). The intensity distribution features an irregular, pseudo-random structure, reminiscent of a speckle pattern with an average grain size of $\lambda / (2 n)$, where $n$ is the refractive index. The characteristic length scale is isotropic, in contrast to the FP-cavity modes that feature a larger transverse than longitudinal wavelength. The wave-vector distribution (Fig.~\ref{fig:passiveModes}D) reveals that the D-cavity mode is a superposition of numerous plane waves in all possible directions. 

These features of the chaotic cavity modes directly affect the lasing dynamics: since the spatial structure of the modes is so fine-grained in all directions, the spatial extent of field intensity variations is too small to create a lensing effect, and additionally there are no dominant propagation directions that light could be focused to. These qualitative differences of the mode structure and the associated length scales compared to FP-cavities result from complex multi-wave interference and can prevent the formation of coherent spatio-temporal structures such as filaments. 

\begin{figure*}[tb]
\begin{center}
\includegraphics[width = 16.0 cm]{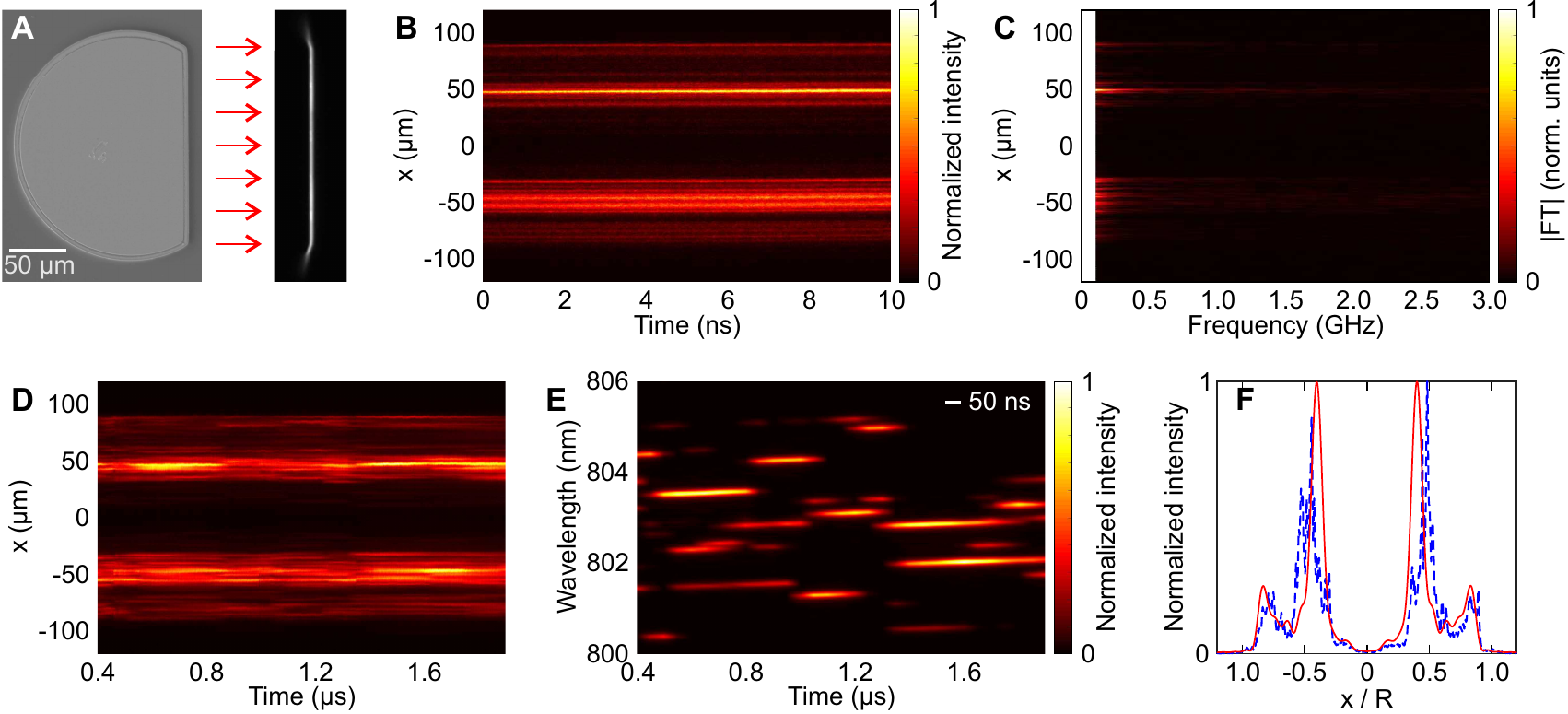}
\end{center}
\caption{Lasing dynamics in a D-cavity with $100~\mu$m radius fabricated by reactive ion etching. 
	(\textbf{A})~Top view SEM image and optical image of the electroluminescence on the straight boundary segment. The pump current for the electroluminescence image was $3$~mA, well below the lasing threshold of $I_{th} = 150$~mA. The intensity profile is homogeneous.
	(\textbf{B})~Spatio-temporal image of the emission intensity, $I(x,t)$, at the straight segment for $500$~mA pump current during a $10$~ns-long interval at $1.4~\mu$s after the start of a 2~$\mu$s-long pump pulse.
	(\textbf{C})~Temporal Fourier transform of $I(x, t)$ in (B), demonstrating the absence of nanosecond-scale oscillations.
	(\textbf{D})~Spatio-temporal image of the emission intensity during the interval $0.4$--$1.9~\mu$s.
	(\textbf{E})~Spectrochronogram for the same pump conditions as in (D), measured with $50$~ns temporal resolution. 
	(\textbf{F})~Lasing emission intensity distribution at the straight segment for $500$~mA pump current, measured with the CCD camera and integrated over a $2~\mu$s-long pulse (blue dashed line), and numerically calculated emission profile of high-$Q$ modes (red solid line).
}
\label{fig:DcavDynamics}
\end{figure*}

We fabricated D-cavity lasers by photolithography and wet or dry (reactive ion) etching [see methods \cite{Methods}]. Figure~\ref{fig:DcavDynamics}A shows a SEM image of a cavity fabricated by reactive ion etching. Figure~\ref{fig:DcavDynamics}B is the spatio-temporal trace of the lasing emission intensity, $I(x,t)$, at the straight segment of the boundary of the D-cavity. Compared to the emission trace in a $10$~ns-long interval for the stripe laser (Fig.~\ref{fig:FPdynamics}A), the D-cavity laser emission has nearly constant intensity and does not exhibit rapid pulsations. The temporal Fourier transform of $I(x,t)$ in Fig.~\ref{fig:DcavDynamics}C confirms the absence of GHz frequency oscillations, in contrast to Fig.~\ref{fig:FPdynamics}C. The spatio-temporal trace of the D-cavity laser (Fig.~\ref{fig:DcavDynamics}D) over a time interval of $1.5~\mu$s reveals temporal fluctuation of the emission intensity on a much longer scale of $\sim 100$~ns. 

The temporal fluctuations of the emission spectrum were measured by a spectrometer equipped with an intensified CCD camera [ICCD, see methods and Fig.~\ref{fig:setup} \cite{Methods}]. The time-resolved emission spectrum (Fig.~\ref{fig:DcavDynamics}E) consists of multiple lasing peaks at any given time. Each peak persists for tens or even hundreds of nanoseconds, and is then replaced by new peaks at different wavelengths. 

To quantify the time scales of the spatio-temporal and spectro-temporal dynamics, we calculated the autocorrelation functions of the spatio- and spectro-temporal data and determined the corresponding correlation times [see Fig.~\ref{fig:ACfunctions} and methods \cite{Methods}]. The correlation times are $\tau_\mathrm{corr}^{(\lambda)} = 94$~ns and $\tau_\mathrm{corr}^{(x)} = 83$~ns, respectively, for the measurements shown in Fig.~\ref{fig:DcavDynamics}. Therefore, the spatio- and spectro-temporal dynamics of the D-cavity laser feature the same characteristic time scales. They are about two orders of magnitude slower than those of the stripe laser ($\leq 1$~ns). These results were further confirmed by measurements of other D-cavity lasers with different size. 

As seen in Figs.~\ref{fig:DcavDynamics}, B and D, the lasing emission from the straight segment of the D-cavity is spatially inhomogeneous. This inhomogeneity is not caused by defects on the sidewall, as confirmed by the smooth electroluminescence profile in Fig.~\ref{fig:DcavDynamics}A. When the pump current increases, a spatially inhomogeneous emission pattern gradually develops (see Fig.~\ref{fig:DcavSWdistrExp}). The intensity profile for $500$~mA, plotted as dashed blue line in Fig.~\ref{fig:DcavDynamics}F, exhibits two distinct length scales. The coarse scale, of the order of several tens of micrometers, represents the size of the dark region in the middle and the bright regions of strong emission around it. The fine scale, of the order of a few micrometers, is the width of the narrow peaks inside the bright regions. Experimentally, the coarse scale is proportional to the cavity size (see Fig.~\ref{fig:DcavSWdistrExp}), while the fine scale is limited by the spatial resolution of the imaging optics. According to numerical simulations [see methods \cite{Methods}], the coarse-scale emission profile reflects the sum of the intensity distributions of the passive D-cavity modes with high quality ($Q$) factors. Those high-$Q$ modes within the gain spectrum correspond to the lasing modes due to their low thresholds, and their intensity distributions determine the total emission profile. The calculated emission intensity profile shown as red solid line in Fig.~\ref{fig:DcavDynamics}F [also see Fig.~\ref{fig:DcavModeSum} \cite{Methods}] agrees well with the coarse structure of the measured emission profile. While the coarse structure is maintained throughout the pulse, the fine-scale peaks appear or disappear over the course of the pulse as different lasing modes turn on or off. 

\begin{figure*}[tb]
\begin{center}
\includegraphics[width = 16.0 cm]{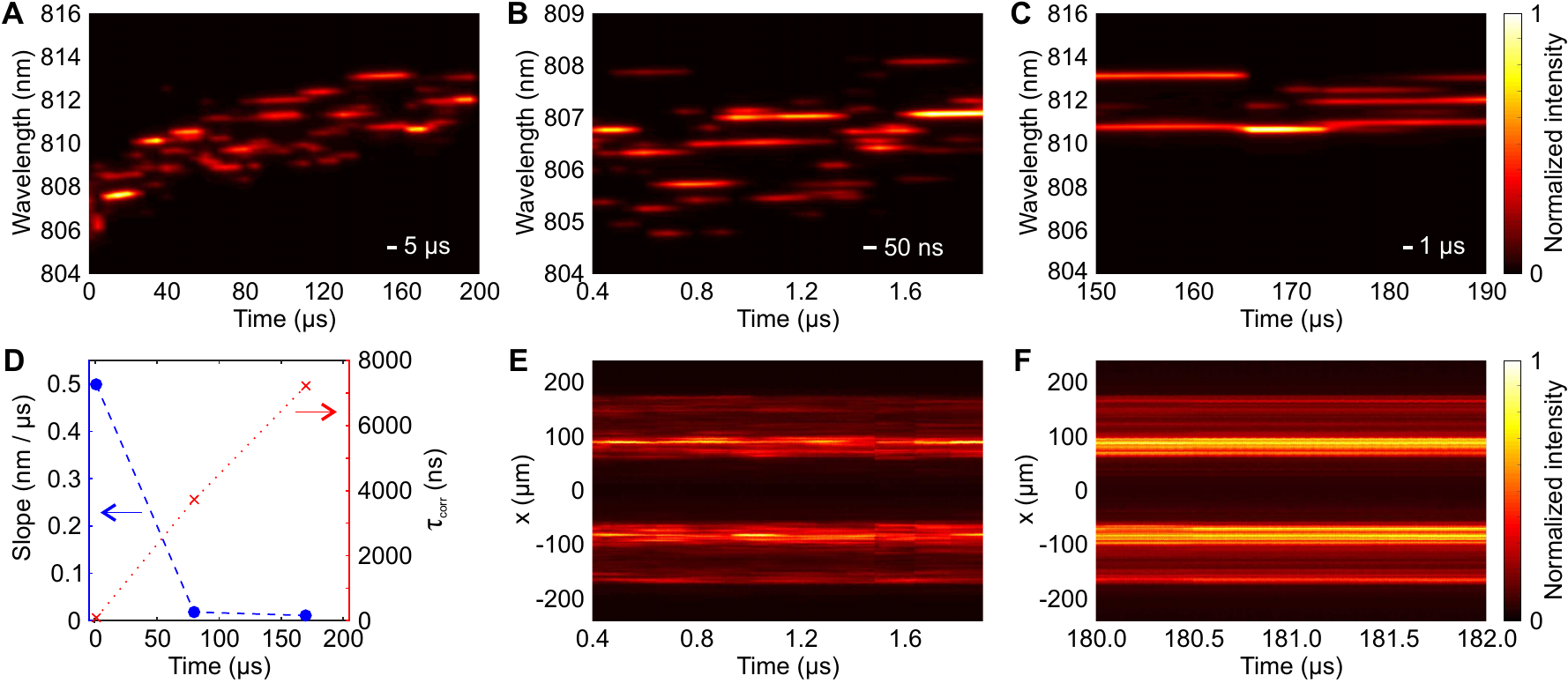}
\end{center}
\caption{Thermal effect on the lasing dynamics. A D-cavity laser fabricated by reactive ion etching with $R = 200~\mu$m radius was pumped by a $t_p = 200~\mu$s pulse. The pump current was $800$~mA, where the lasing threshold was $I_{th} = 300$~mA.
(\textbf{A}) Spectrochronogram of the lasing emission for $0$--$200~\mu$s measured with $5~\mu$s temporal resolution. The spectral shift to longer wavelengths is caused by an increase of the sample temperature.
(\textbf{B}) Spectrochronogram for $0.4$--$1.9~\mu$s measured with $50$~ns temporal resolution.
(\textbf{C}) Spectrochronogram for $150$--$190~\mu$s measured with $1~\mu$s temporal resolution.
(\textbf{D}) Rate of the red-shift of the center of mass of the emission spectra (blue circles) and the spectral correlation times $\tau_\mathrm{corr}^{(\lambda)}$ (red crosses) at different times during the $200~\mu$s pulse. The red-shift slope decreases by almost two orders of magnitude as the sample  temperature stabilizes, and conversely the spectral correlation time increases by two orders of magnitude. 
(\textbf{E}) Spatio-temporal image of the lasing emission during $0.4$--$1.9~\mu$s and (\textbf{F}) during $180$--$182~\mu$s, showing the spatio-temporal dynamics becomes more stable with time.
}
\label{fig:DcavDynamicsLong}
\end{figure*}

Next we show that the remaining fluctuations of the laser emission from wave-chaotic cavities result from thermal effects. The current injection causes sample heating, which modifies the refractive index of the cavity and the gain spectrum of the quantum well. Consequently, the lasing modes may change, leading to dynamic variations of the emission spectra as well as the spatial emission intensity distributions. In order to investigate the thermal effects, we increased the pump pulse length $t_p$ to 200~$\mu$s. After the turn-on of the pump current, the sample temperature first rose quickly, then gradually stabilized. If heating effects were relevant, the lasing dynamics would slow down over time. 

Figure~\ref{fig:DcavDynamicsLong}A presents the spectro-temporal data for a D-cavity laser with $R = 200~\mu$m. Over the time interval of $t_p = 200~\mu$s, the lasing spectrum exhibits a continuous shift to longer wavelengths due to the increase of the sample temperature. However, the red shift of the lasing spectrum notably slows down during the later part of the pump pulse, and individual peaks last longer in time. We computed the center of mass (COM) for the time-resolved spectrum $\lambda_\mathrm{COM}(t)$, and found it is fitted well by an exponential function $\lambda_\mathrm{COM}^\mathrm{(fit)}(t) = \lambda_0 - \lambda_1 \exp(-t / \tau_{th})$, with the decay time $\tau_{th} = 174~\mu$s [see methods \cite{Methods}]. The slope $d\lambda_\mathrm{COM}^\mathrm{(fit)}(t)/dt$ gives the rate of the spectral shift. The sample temperature gradually stabilizes during the pulse as indicated by the decreasing slope of $\lambda_\mathrm{COM}(t)$ from $0.5~\mathrm{nm}/\mu$s during the first two microseconds to $0.01~\mathrm{nm}/\mu$s at $170~\mu$s (see Fig.~\ref{fig:DcavDynamicsLong}D). 

To characterize the change of the time scale of the emission fluctuations, we measured the time-resolved spectra at different times during the $200~\mu$s pulse with better temporal resolution. The spectral correlation time for a D-cavity laser increases from $\tau_\mathrm{corr}^{(\lambda)} = 90$~ns during the first $2~\mu$s (Fig.~\ref{fig:DcavDynamicsLong}B) to $7.2~\mu$s during $150$--$190~\mu$s (Fig.~\ref{fig:DcavDynamicsLong}C). Figure~\ref{fig:DcavDynamicsLong}D shows the correlation times and slope of $\lambda_\mathrm{COM}$ at different times during the pulse, illustrating how the emission fluctuations slow down as the temperature stabilizes. Spatio-temporal measurements also confirmed the lasing dynamics become more stable with time (Fig.~\ref{fig:DcavDynamicsLong}, E and F). 

These results illustrate the effect of the temperature change on the lasing dynamics, and indicate that better thermal management can lead to a further stabilization of the temporal dynamics of wave-chaotic lasers. This is in stark contrast to the wide stripe lasers which did not exhibit a stable dynamics at all. Fast oscillations and pulsations on a nanosecond time scale persisted until $200~\mu$s, even though the emission spectra indicated the sample had reached a thermal equilibrium after $\sim 50~\mu$s [Fig.~\ref{fig:FPlongPulse} \cite{Methods}]. 

We also tested the D-cavity lasers fabricated by wet chemical etching. Although the cavity sidewalls are not vertical and rougher than for fabrication by reactive ion etching, the spatio- and spectro-temporal dynamics of the lasing emission is very similar [Fig.~\ref{fig:DcavDynamicsWet} \cite{Methods}]. These results demonstrate the robustness of the stable lasing dynamics in a wave-chaotic cavity against fabrication imperfections. However, the spatial emission profile differs from that of a dry-etched cavity. This is attributed to the modification of the mode structures by the rough boundary, and confirmed by numerical simulations [Fig.~\ref{fig:DcavRoughModes} \cite{Methods}]. Even in the presence of boundary roughness, the complex wave interference persists in the wave-chaotic cavities and suppresses the formation of filamentation and spatio-temporal instabilities. Consequently the lasing emission profile is dictated by the passive cavity mode structure.

\section*{Lasing dynamics in disordered cavities}

While the wave-chaotic cavities can efficiently suppress lasing instabilities, they lack emission directionality due to the absence of a predominant propagation direction. Therefore the question arises if we can suppress lasing instabilities via complex wave interference while having directional emission. 

\begin{figure*}[tb]
\begin{center}
\includegraphics[width = 15.5 cm]{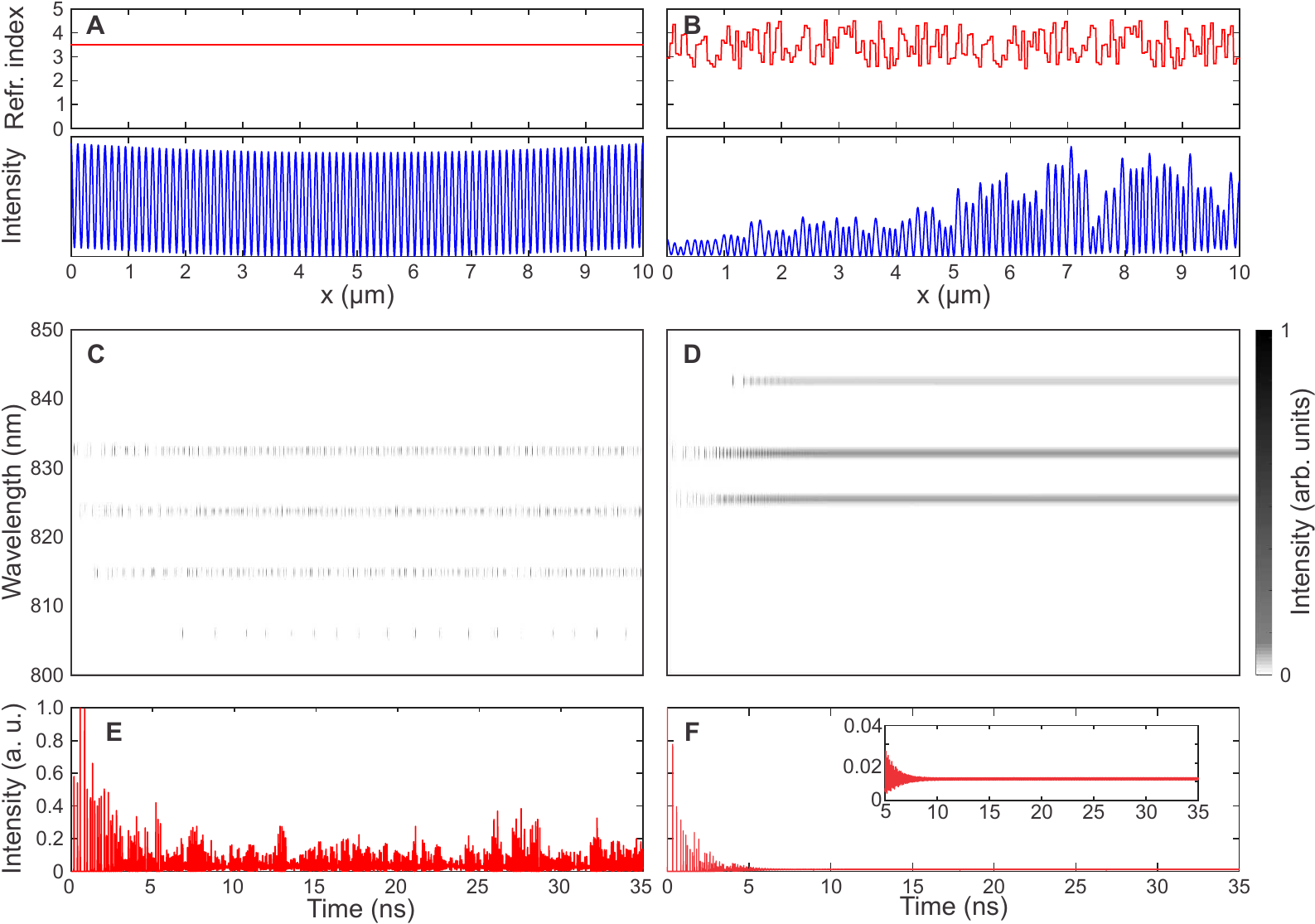}
\end{center}
\caption{Simulation of the lasing dynamics in one-dimensional cavities with homogeneous refractive index profile and spatially-varying index profile, respectively, at the same pump current density $J = 500$~A~cm$^{-2}$.
(\textbf{A, B}) Spatial distribution of the refractive index (red line) and the field intensity (blue line) for a mode at $\lambda = 833.3$~nm in the homogeneous cavity (A) and for a mode at $\lambda = 833.8$~nm in the disordered cavity (B). 
(\textbf{C, D}) Spectrochronogram of the emission intensity from the homogeneous cavity (C) and from the disordered cavity (D).
(\textbf{E, F}) Total output intensity for the homogeneous cavity (E) and for the disordered cavity (F).
}
\label{fig:activeModes}
\end{figure*}

We consider a simple 1D dielectric slab cavity with random fluctuations of the refractive index (Fig.~\ref{fig:activeModes}B). The index fluctuations generate multiple reflected waves that interfere subsequently. Thus, the resonant modes no longer have a constant frequency spacing and their spatial profiles become irregular with varying spatial scales (Figs.~\ref{fig:activeModes}B and \ref{fig:randCavModes}), reminiscent of the modes in a 2D wave-chaotic cavity (Fig.~\ref{fig:DcavModesCuts}). 

To simulate their lasing dynamics, we solved the semiconductor Maxwell-Bloch equations in the time domain. Our full-wave model goes beyond the slowly varying envelope / rotating wave (in time) and paraxial (in space) approximations, fully resolving the spatio-temporal dynamics on sub-optical cycle and sub-wavelength scales [see methods \cite{Methods}]. The population inversion-dependent optical gain has an asymmetric spectrum, which closely reproduces that of a semiconductor quantum well. Taking into account the dynamical coupling between the light field and the carrier system, we include all spatio-temporal and nonlinear effects such as spatial hole burning and multiple wave mixing mediated by the carriers~\cite{Guazzotti2016,Boehringer2008Both}. 

We compare the simulated lasing dynamics of a disordered cavity to that of a homogeneous cavity with regular mode structure in Fig.~\ref{fig:activeModes}A. The disordered cavity features stable lasing dynamics over a wide range of pump currents, while lasing in the homogeneous cavity is stable only just above threshold and becomes unstable with increasing pump current [Figs.~\ref{fig:ActiveSimHomogeneous} and \ref{fig:ActiveSimDisordered} \cite{Methods}]. For example, when the pump current is about five times of the threshold ($J_{th} = 104~\mathrm{A}~\mathrm{cm}^{-2}$), four longitudinal modes lase in the homogeneous cavity, and all modes pulsate irregularly on a sub-nanosecond time scale (Fig.~\ref{fig:activeModes}C). The total emission intensity in Fig.~\ref{fig:activeModes}E fluctuates in time, and does not approach a constant value even well beyond the transient dynamics. These instabilities are caused by the nonlinear interactions between the lasing modes and the gain medium through processes such as spatial hole burning and multi-wave-mixing \cite{Furfaro2004}. 

The disordered cavity with almost identical lasing threshold ($J_{th} = 96~\mathrm{A}~\mathrm{cm}^{-2}$) as the homogeneous cavity has three modes lasing at the same pump current density. After some initial pulsations, each lasing mode reaches a steady state (Fig.~\ref{fig:activeModes}D). The total output intensity also approaches a constant value beyond the transient period (Fig.~\ref{fig:activeModes}F). The stable state of multimode lasing sets in faster at higher pump current [Fig.~\ref{fig:ActiveSimDisordered} \cite{Methods}]. 

Therefore, even in a 1D cavity, the interference of multiple scattered waves with random phases can lead to stable lasing dynamics, and the stabilization is complete in the absence of thermal effects. These results confirm the generic nature of our scheme to suppress spatio-temporal instabilities by increasing the spatial complexity of the lasing modes. 

\section*{Discussion and conclusion}

Our approach for obtaining a stable state of multimode lasing in broad-area edge-emitting semiconductor lasers is fundamentally different from previous ones in several respects. Most previous approaches aim at suppressing the spatio-temporal instabilities and the formation of self-organized structures like filaments by minimizing the number of lasing modes. Our approach maintains multimode lasing while achieving stable temporal dynamics by tailoring the spatial properties of the lasing modes using resonators with chaotic ray dynamics or with random refractive index fluctuations. Although the mechanisms causing lasing instabilities in 1D and 2D cavities are different, both are disrupted by complex wave interference. Since this process is present across the whole cavity, we attain global suppression of the instabilities, in contrast to schemes like injection and feedback that can influence the dynamics only locally. 

It is important to note that our scheme of achieving stable multimode operation is very robust with respect to perturbations such as boundary roughness, since they do not qualitatively change the already pseudo-random structure of the lasing modes. Although small modifications of the cavity geometry of broad-area edge-emitting semiconductor lasers were considered previously \cite{Adachihara1993,Levy1997,Simmendinger1999,Buettner2005}, a dominant propagation direction and thus well-defined wave fronts were maintained, and the spatial scales of the modes were not significantly modified, in stark contrast to the wave-chaotic and disordered cavities presented here. 

Although the multimode operation of D-cavity lasers produces emission with relatively low spatial coherence \cite{Redding2015}, which prevents tight focusing, the temporal stability of the lasing power and the emission profile, as shown in this work, is essential to produce stable beam profiles necessary for many high-power applications. For example, laser processing of materials and devices requires diverse beam shapes such as circular flat-top, square, rectangle or line profiles, and various beam-shaping techniques have been developed in recent years \cite{Fuse2015}. Low spatial coherence of the laser beams prevents coherent artifacts and enables smooth intensity profiles, e.g., the D-cavity laser emission may be coupled to a multimode fiber to produce a stable flat-top beam free of speckle. Another potential application is pumping high-power multimode fiber lasers and amplifiers. 

In previous studies, broad-area VCSELs with pulsed pumping demonstrated non-modal emission with low spatial coherence, when the interplay between a rapid thermal chirp and the build-up of a thermal lens breaks up the global cavity modes \cite{Mandre2008}. As the VCSEL becomes thermally stable with time, the multimode operation resumes and fast temporal dynamics appears. This is fundamentally different from the wave-chaotic cavities in which the stable state of lasing is maintained in multimode operation. It should be mentioned that random fiber lasers can also exhibit temporal fluctuations \cite{Turitsyn2010}, which disappear for stronger pumping. Both the mechanism inducing the instabilities (interplay between stimulated Brillouin scattering (SBS) and Rayleigh scattering) and that quenching the instabilities (suppression of SBS) are distinct from those for the 1D disordered semiconductor lasers we simulated [see supplementary materials \cite{Methods}]. 

We therefore propose the demonstrated suppression of lasing instabilities by means of complex multi-wave interference as a new paradigm for manipulating the temporal dynamics of multimode lasers. We believe it is generally applicable to other high-power lasers exhibiting instabilities such as broad-area VCSELs and solid-state lasers, as well as multimode fiber lasers and amplifiers. By deforming the cavity or fiber cross section or introducing random refractive index fluctuations, the spatial mode structure becomes speckled, preventing lens formation and self-focusing instabilities. On a more general level, this work opens a new direction of research combining concepts from both wave-dynamical chaos and deterministic chaos. This combination and its implications have been barely investigated so far in lasers or other nonlinear wave-dynamical systems. We expect the idea of manipulating nonlinear temporal dynamics by disrupting the formation of self-organized structures will have a significant impact not only on laser physics but will find applications in other systems with complex spatio-temporal dynamics as well. 


\section*{Acknowledgments}
H.C.\ and S.B.\ thank A.\ Douglas Stone, Hakan T\"ureci, Li Ge, Jonathan Andreasen and Christian Vanneste for fruitful discussions.

\textit{Funding.}
The work conducted at Yale University is supported partly by the Office of Naval Research (ONR) with a MURI grant N00014-13-1-0649, and by the Air Force Office of Scientific Research (AFOSR) under grant no.\ FA9550-16-1-0416.
The research at Imperial College London is partly supported by the Engineering and Physical Sciences Research Council (EPSRC) UK through the projects EP/L024926/1 and EP/L027151/1, by the AFOSR under grant no.\ FA9550-14-1-0181, and with funding by the European Regional Development Fund through the Welsh Government.
For the work at Nanyang Technological University, funding support is acknowledged from the Singapore Ministry of Education Tier 2 Program (MOE 2016-T2-1-128) and the Singapore National Research Foundation Competitive Research Program (NRF-CRP18-2017-02).

\textit{Author contributions.}
S.B.\ conducted experimental characterization and data analysis, and prepared the manuscript.
H.Y., Y.Z., X.H.\ and Q.W.\ did sample fabrications.
K.K.\ calculated the resonant modes in D-cavities.  
S.G.\ and S.S.O.\ modeled the semiconductor laser dynamics.
O.H.\ guided the theory of complex semiconductor laser dynamics and edited the paper.
H.C.\ proposed the idea, initiated the research and edited the paper.

\textit{Competing interests.}
The authors declare no competing financial interests.

\textit{Data and materials availability.}
All data needed to evaluate the conclusions in the paper are present in the paper or the supplementary materials.


\renewcommand{\theequation}{S\arabic{equation}}
\renewcommand{\thefigure}{S\arabic{figure}}
\setcounter{figure}{0}
\setcounter{equation}{0}

\section*{Materials and methods}

\subsection*{Sample fabrication}

\textit{Epitaxial wafer}. The semiconductor lasers were fabricated with a commercial diode laser wafer (Q-Photonics QEWLD-808). The epitaxial structure grown on an n-doped GaAs wafer consists (from bottom to top) of a $1.4~\mu$m-thick n-doped Al$_{0.55}$Ga$_{0.45}$As cladding layer, an undoped $400$~nm-thick Al$_{0.37}$Ga$_{0.63}$As guiding layer with a $12$~nm-thick GaAs quantum well in the middle, and a $1.5~\mu$m-thick p-doped Al$_{0.55}$Ga$_{0.45}$As cladding layer capped by a $500$~nm-thick GaAs layer. 

\textit{Reactive ion etching}. First, the backside of the wafer was metalized with Ni/Ge/Au/Ni/Au layers (thicknesses $5/25/100/20/200$~nm, respectively) and thermally annealed at $385^\circ$C for $30$~s. A layer of $300$~nm SiO$_2$ was then deposited on the front side by plasma enhanced chemical vapor deposition (PECVD). The cavity shapes were defined by photolithography, and were then transferred to the SiO$_2$ layer by reactive ion etching (RIE) with an O$_2$ ($30$~sccm) and CF$_4$ ($30$~sccm) gas mixture. After the removal of the photoresist on top of the SiO$_2$, the D-shaped cavities were formed by inductively coupled plasma (ICP) dry etching of GaAs/AlGaAs with SiO$_2$ as hard mask. A BCl$_3$/Cl$_2$/Ar plasma mixture with flow rates of $4.5/4.0/5.0$~sccm was used to etch all the way through the guiding layer and partially into the bottom cladding layer. The deep etch (about $3.1~\mu$m) ensures sufficient index contrast and optical reflection at the sidewalls of the disks. After the ICP dry etching, the SiO$_2$ hard mask was removed by the RIE process. A negative resist photolithography was conducted to define the top metal contacts, followed by Ti/Au (thicknesses $20/200$~nm) metal deposition and lift-off. The top metal contacts are slightly smaller than the cavities to facilitate the alignment in the second photolithography. Finally, an O$_2$ plasma etching process was used to ensure the cleanness of the D-cavity sidewalls. 

\textit{Wet chemical etching}. After the metal deposition on the backside of the wafer, the cavity shapes were defined on the front side by photolithography. The top metal contacts consisting of Ti/Au layers (thicknesses $20/300$~nm) were deposited. After the liftoff, the top metal contacts were used as masks for the wet etching process going all the way through the guiding layer and partially into the bottom cladding layer using a H$_3$PO$_4$:H$_2$O$_2$:H$_2$O solution. The etch depth was $3.1~\mu$m.

\begin{figure}[tb]
\begin{center}
\includegraphics[width = 8.4 cm]{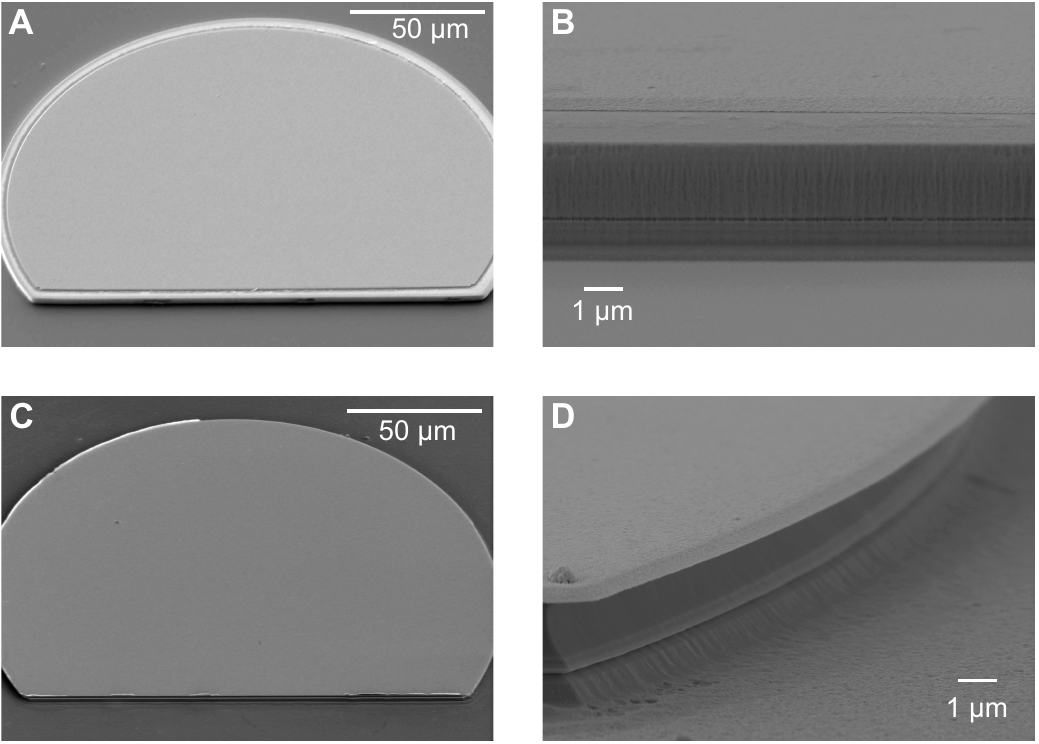}
\end{center}
\caption{SEM images of D-cavities with radius $R = 100~\mu$m fabricated by \textbf{(A, B)} reactive ion and \textbf{(C, D)} wet etching. \textbf{(A, C)}~View of the entire cavities with a perspective angle of $45^\circ$ with respect to the surface normal. \textbf{(B, D)}~Magnification of the cavity sidewall with a perspective angle of $80^\circ$.}
\label{fig:SEM-Dcav}
\end{figure}

\textit{SEM images}. Figures~\ref{fig:SEM-Dcav}, A and B, show the SEM images of D-cavities fabricated by reactive ion etching, and Figs.~\ref{fig:SEM-Dcav}, C and D, cavities fabricated by wet chemical etching. Figure~\ref{fig:SEM-Dcav}A shows that the top metal contact does not fully extend to the boundary of the dry-etched cavity. In Fig.~\ref{fig:SEM-Dcav}D, the sidewalls of the wet-etched D-cavities are slightly sloped and have more roughness than the dry-etched cavities. Since the top metal contacts are used as etch mask they are undercut in the wet etching process. 

\textit{Stripe laser}. The conventional stripe lasers were fabricated from the same wafer in a similar procedure but without etching. The top metal contact has the shape of a stripe that is $50$--$100~\mu$m wide. After depositing the top metal contact, the wafer was cleaved into about $1$~mm-long pieces, and the reflections from two cleaved facets (without coating) form the Fabry-Perot cavities. The current injection from the stripe contact provides gain guiding in the direction perpendicular to the stripe axis.

\subsection*{Optical measurements}

\textit{Electrical pumping}. The electrical current was injected through a tungsten needle to the top metal contact, and a copper plate, on which the sample was mounted, served as bottom contact. The diode driver (DEI Scientific, PCX-7401) produced rectangular current pulses of length $t_p = 2$--$200~\mu$s. The repetition rate of $f_\mathrm{rep} = 9$~Hz reduced the heating effect due to the low duty cycle. 

\begin{figure}[tb]
\begin{center}
\includegraphics[width = 8.4 cm]{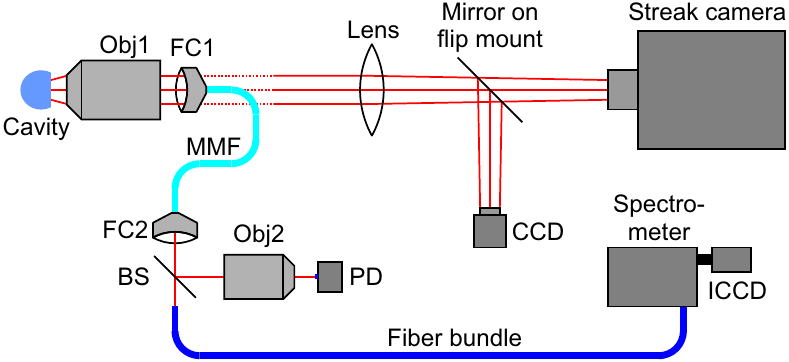}
\end{center}
\caption{Schematic of the experimental setup for the spatio- and spectro-temporal measurements of lasing dynamics. The edge emission from the laser cavity is collected by an objective (Obj1) and fed to a multimode fiber (MMF) via a fiber collimator (FC1). The output from the fiber, after being collimated by the second fiber collimator (FC2), is split by a beam splitter (BS), and in one arm focused on a fast photo-diode (PD) by an objective (Obj2) and in the other arm transmitted to a spectrometer via a fiber bundle. The time-resolved spectra are measured by an intensified CCD camera (ICCD). For spatial and spatio-temporal measurements, FC1 is removed and the emission intensity distribution on the cavity boundary is imaged onto a CCD camera or a streak camera via the objective and a lens in $(2f_1 - 2f_2)$ configuration.}
\label{fig:setup}
\end{figure}

\textit{Experimental setup}. A schematic of the optical setup is presented in Fig.~\ref{fig:setup}. The emission from the edge of the cavity was collected by either a $10\times$ microscope objective ($\mathrm{NA} = 0.25$) or a $20\times$ objective ($\mathrm{NA} = 0.40$), depending on the cavity size, and imaged on a CCD camera (Allied Vision Mako G-125B) for time-integrated measurement or on the entrance slit ($100~\mu$m slit width) of a streak camera (Hamamatsu C5680) for time-resolved measurement.

\textit{Time-resolved measurements}. The streak camera is equipped with a fast sweep unit (M5676) that records the spatio-temporal traces in a time window of length $t_r \leq 50$~ns, with the temporal resolution at least $200$ times smaller than $t_r$. Alternatively, a fiber collimator ($\mathrm{NA} = 0.50$) was placed behind the objective to feed the emission into a multimode fiber (diameter $600~\mu$m, $\mathrm{NA} = 0.48$). From there, the emission was focused by another microscope objective ($10\times$, $\mathrm{NA} = 0.25$) onto a fast photo-diode (Alphalas UPD-200-SP, rise time $< 175$~ps), which was connected to an oscilloscope (Tektronix MDO3104, $1$~GHz bandwidth) for measuring the total emission intensity as a function of time. 

The emission collected by the multimode fiber was transmitted to a fiber bundle connected to an imaging monochromator (Acton SP300i) equipped with an intensified CCD (ICCD) camera (Andor iStar DH312T-18U-73) for the measurement of time-resolved emission spectra. The triggering and gating of the various instruments was controlled by a computer-controlled digital delay generator built into the ICCD, which determined the position of the active time window of the streak camera and ICCD, respectively, during the emission pulse. 

The initial measurements by the fast photo-diode allowed to observe the temporal dynamics of the entire emission pulse with relatively low temporal resolution (bandwidth $1$~GHz). In order to investigate the dynamics with higher temporal resolution, the streak camera and ICCD were used to measure the emission during a short time window of a single emission pulse. The lasers were then pumped repeatedly with the same conditions, and the time window of measurement was moved step by step via the delay generator to scan the whole emission pulse (or at least a larger part of it). All measurements with the streak camera and the ICCD are in single-shot mode. 

\begin{figure*}[tb]
\begin{center}
\includegraphics[width = 13.0 cm]{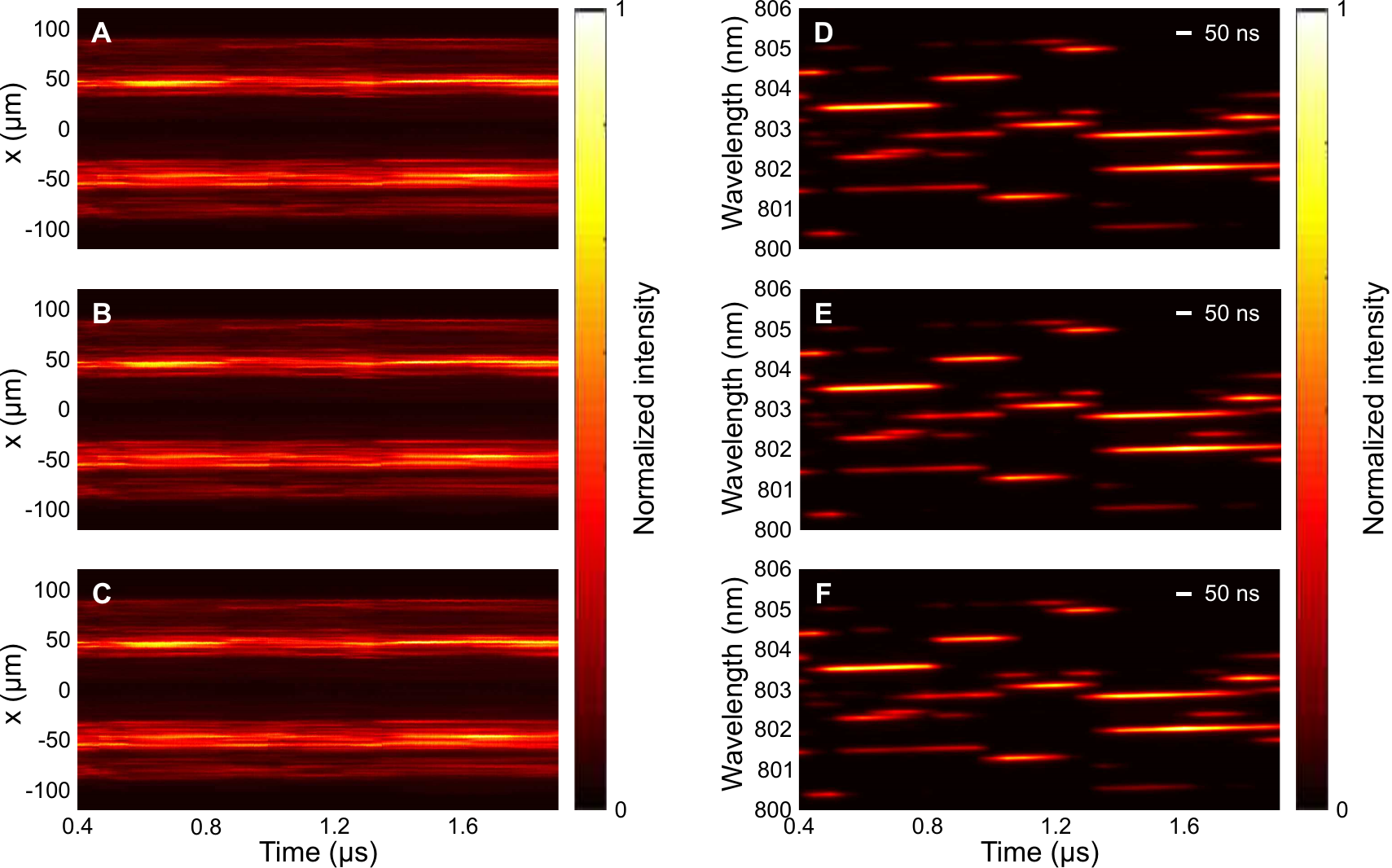}
\end{center}
\caption{Lasing dynamics of a D-cavity fabricated by reactive ion etching with $100~\mu$m radius for a pump current of $500$~mA in the interval $0.4$--$1.9~\mu$s after the start of a 2~$\mu$s-long pump pulse (cf.\ Fig.~\ref{fig:DcavDynamics}). The left column (\textbf{A--C}) shows the spatio-temporal images of three consecutive pump pulses, the right column (\textbf{D--F}) the corresponding spectrochronograms measured with $50$~ns temporal resolution.}
\label{fig:D100Repeat}
\end{figure*}

\textit{Repeatability}. It is important to note that the lasing dynamics was repeatable from pulse to pulse with high precision. In order to verify the stability of the system and the deterministic nature of the temporal dynamics, all spatio- and spectro-temporal measurements were repeated three times with consecutive current pulses $p = 1,2,3$. Figure~\ref{fig:D100Repeat} shows the comparison of the data acquired from consecutive pulses for a D-cavity with $100~\mu$m radius. It is the same data set as that shown in Fig.~\ref{fig:DcavDynamics}. Both the spatio-temporal images in Fig.~\ref{fig:D100Repeat}, A--C, and the spectrochronograms in Fig.~\ref{fig:D100Repeat}, D--F, show very good agreement with one another down to even small details. The good repeatability demonstrates that the observed dynamics is deterministic in nature, whereas stochastic effects have no observable influence. The spectro-temporal and spatio-temporal measurements of individual pulses could thus be pieced together to obtain the time evolution of spectra and spatial intensity distributions during the entire pulse as presented. Repeated measurements with cavities of different sizes and different realizations of surface roughness yielded qualitatively the same results. 

\textit{Spatio-spectral measurements}. The spectra of D-cavity lasers were also measured with spatial resolution. The straight segment of the boundary of the D-cavity was imaged by a $10\times$ microscope objective ($\mathrm{NA} = 0.25$) onto a line-to-line fiber bundle (Thorlabs BFA200LS02), which consists of seven fibers ($200~\mu$m diameter and $\mathrm{NA} = 0.22$). The magnification was chosen such that the image of the D-cavity sidewall covers the input facets of all seven fibers arranged in a line. The output end of the fiber bundle was imaged onto the entrance slit of an imaging spectrometer (Acton SP300i) with the line of seven fibers parallel to the slit. In this way, the emission spectra from  seven different sections of the D-cavity sidewall were measured by the ICCD camera mounted to the output port of the imaging spectrometer. 

\subsection*{Autocorrelation functions and correlation times}

The time scales of the spatio-temporal and spectro-temporal dynamics are determined from the autocorrelation (AC) functions 
\begin{equation} \label{eq:ACfunDef} C(\tau) = \sum \limits_r \left< I_\mathrm{fluc}(t, r) I_\mathrm{fluc}(t + \tau, r) \right>_t \end{equation}
for the fluctuating part of the emission intensity $I_\mathrm{fluc}(t, r) = [I(t, r) - \left< I(t, r) \right>_t] / \sigma_I(r) $, where $r$ is either the wavelength $\lambda$ or the spatial position $x$, and $\sigma_I(r)$ is the standard deviation of $I(t, r)$ for a given $r$. The AC functions are normalized to $C(0) = 1$, and their widths yield the correlation times $\tau_\mathrm{corr}^{(r)}$. 

\begin{figure*}[tb]
\begin{center}
\includegraphics[width = 11.0 cm]{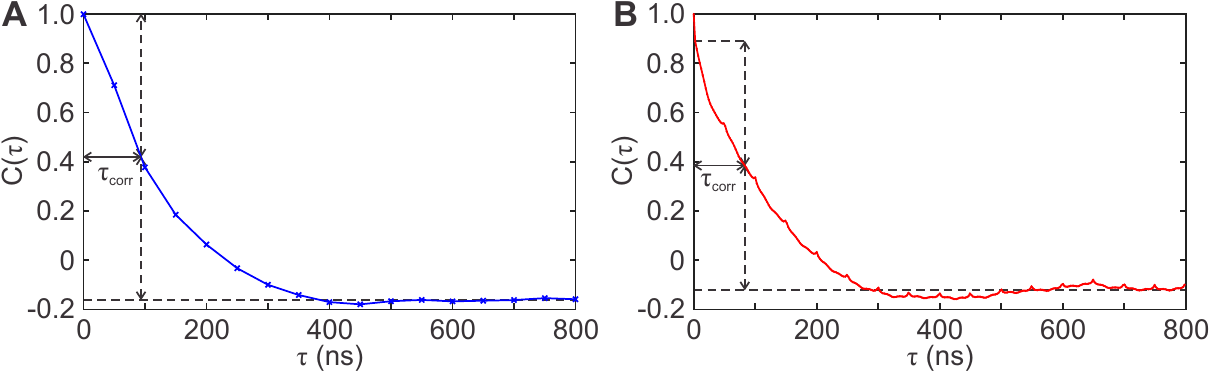}
\end{center}
\caption{\textbf{(A)} Spectral and \textbf{(B)} spatial autocorrelation functions corresponding to the data shown in Fig.~\ref{fig:DcavDynamics}. The bottom horizontal dashed lines indicate the base values of the AC functions, and the top horizontal dashed line in (B) indicates the maximum value (after the initial sharp drop of the AC function) used for determining the half width at half maximum (HWHM) of the AC function. The correlation times, given by the HWHM, are $\tau_\mathrm{corr}^{(\lambda)} = 94$~ns and $\tau_\mathrm{corr}^{(x)} = 83$~ns, respectively, denoted by the horizontal (solid black) line with double-sided arrows.}
\label{fig:ACfunctions}
\end{figure*}

It is important to take into account the inevitable noise in the measured data $I(t, r)$ when calculating the AC functions and the correlation times. First, the sum over $r$ in Eq.~(\ref{eq:ACfunDef}) only includes spatial positions $x$ or emission wavelengths $\lambda$ for which the time average $\left<I(t, r)\right>_t$ is above the noise floor. In addition, the streak images were binned over $50$ time pixels (corresponding to about $2.5$~ns) before computing the spatial correlation functions. The spectral AC function corresponding to the spectrochronogram in Fig.~\ref{fig:DcavDynamics}E is shown in Fig.~\ref{fig:ACfunctions}A, and the spatial AC function corresponding to the spatio-temporal image in Fig.~\ref{fig:DcavDynamics}D is shown in Fig.~\ref{fig:ACfunctions}B. 

The correlation time $\tau_\mathrm{corr}$ is defined as the half width at half maximum (HWHM) of the AC function, marked by the horizontal (solid black) line with double-sided arrows, with respect to the base value of the AC function, indicated by the lower horizontal dashed lines in Figs.~\ref{fig:ACfunctions}, A and B. In the case of the spatial autocorrelation function in Fig.~\ref{fig:ACfunctions}B, however, the measurement noise leads to a sharp drop of $C(\tau)$ as soon as $\tau$ deviates from $0$. A further increase of $\tau$ leads to a gradual decay of $C$. Hence, the value of $C$ at $\tau = 0$ is not used as the maximum for determining $\tau_\mathrm{corr}^{(x)}$, but instead the value of $C(\tau)$ after the initial drop (indicated by the top horizontal dashed line in Fig.~\ref{fig:ACfunctions}B) was used as the maximum for determining the HWHM. 

\subsection*{Time evolution of center of mass of lasing spectra}

The sample gradually heats up during the pump current pulse, which leads to changes in the cavity resonant modes and the gain spectrum of the quantum well. It should be noted that the sample was mounted on a copper block which acted as a large heat sink, and all experiments were conducted at ambient temperature. The heating leads to a notable red shift of the emission spectra during the pump pulse as shown in Fig.~\ref{fig:DcavDynamicsLong}. The effect of the heating is quantified by calculating the center of mass (COM) of the lasing spectra, $\lambda_\mathrm{COM}(t) = \int d\lambda \, I(\lambda, t) \lambda / [\int d\lambda \, I(\lambda, t)]$. 

For the $2~\mu$s-long current pulse shown in Fig.~\ref{fig:DcavDynamicsLong}B, the increase of the COM is well described by a linear function with a slope of $0.50~\textrm{nm} / \mu$s. For the $200~\mu$s long pulse shown in Fig.~\ref{fig:DcavDynamicsLong}A, the COM keeps increasing but the slope decreases with time. An exponential function $\lambda_\mathrm{COM}^\mathrm{(fit)}(t) = \lambda_0 - \lambda_1 \exp(-t / \tau_{th})$ fits the COM well in the time interval of $40$--$200~\mu$s with a time constant of $\tau_{th} = 174~\mu$s. The corresponding slope, calculated as the derivative of the fit function, decreases from about $0.02~\textrm{nm} / \mu$s at $40~\mu$s to less than $0.01~\textrm{nm} / \mu$s at $200~\mu$s. This value is almost two orders of magnitude smaller than that during the first two microseconds, demonstrating the gradual stabilization of the sample temperature during the long pulse, even though thermal equilibrium is not yet reached. This leads to a slow-down of the lasing dynamics as shown in Fig.~\ref{fig:DcavDynamicsLong}. 

\subsection*{Passive cavity mode calculations}

\textit{Fabry-Perot cavity}. The classical ray dynamics in stripe lasers with rectangular Fabry-Perot cavities is regular and the geometry is separable. Hence the quantization conditions in longitudinal and transverse direction are independent, yielding the longitudinal wave number $k_l = n_l \pi / L$ and the transverse wave number $k_t = n_t \pi / W$, respectively, where $L$ is the stripe length and $W$ the effective width. The quantum numbers $n_l$ and $n_t$ are the number of antinodes in the longitudinal and transverse directions of the resonance field distribution \cite{StoeckmannBuch2000}. The wave number of the resonance is then given by $k = 2 \pi / \lambda = [k_t^2 + k_l^2]^{1/2} / n$, where $n$ is the refractive index. Since the light field propagates predominantly in the longitudinal direction, $k_l \gg k_t$, the transverse wavelength $\lambda_t = 2 \pi / k_t$ is much longer than the longitudinal wavelength $\lambda_l = 2 \pi / k_l$. For a typical $L = 1$~mm-long cavity, the longitudinal quantum number will be of the order of $n_l \simeq 8400$ with a corresponding longitudinal wavelength $\lambda_l$ only marginally larger than $\lambda / n$. In contrast, the transverse quantum number is of the order of $n_t = 1$--$10$, yielding a transverse wavelength of the order of at least several micrometers. 

\textit{D-cavities}. We calculated the modes of a passive D-cavity with radius $R = 20~\mu$m and refractive index $n = 3.37$ with COMSOL. The simulations were made for transverse electric (TE) polarization (electric field parallel to the cavity plane) since the lasing modes are TE polarized due to preferential gain of the semiconductor quantum well. These cavity resonances (also called quasi-bound modes) are the solutions of the scalar Helmholtz equation
\begin{equation} [\Delta + n^2(x, y) k^2] H_z(x, y) = 0 \end{equation}
with outgoing wave boundary conditions where $H_z$ is the vertical component of the magnetic field. Examples of the calculated modes are shown in Figs.~\ref{fig:passiveModes}C and \ref{fig:DcavModesCuts}. 

\textit{Surface roughness}. To investigate the effect of surface roughness on the modes, the boundary of a D-cavity with radius $R = 10~\mu$m was perturbed by adding a random superposition of high-order harmonics. Along the circular part of the cavity, the local radius was modified as
\begin{equation} r(\varphi) = R + \sum \limits_{m = m_1}^{m_2} a_m \cos(m \varphi + \theta_m) \end{equation}
where $\varphi$ is the azimuthal angle, and the perturbation amplitudes $a_m$ and the phases $\theta_m$ are random variables in the range of $a_m \in [-25, 25]$~nm and $\theta_m \in [0, 2 \pi]$, respectively. Along the straight segment, the local cavity boundary was determined by
\begin{equation} y(x) = R / 2 + \sum \limits_{m = m_1}^{m_2} a_m' \cos(m x / R + \theta_m') \end{equation}
with $a_m'$ and $\theta_m'$ random variables like $a_m$ and $\theta_m$. The range of the harmonics was from $m_1 = 5$ to $m_2 = 42$, where $m_1$ was chosen such that the maximal length scale of the surface roughness was about $2~\mu$m in agreement with the SEM images, and $m_2$ such that the minimal length scale was $\lambda / n$ since any features smaller than the wavelength are not really resolved by the electromagnetic fields.

\textit{Disordered cavity}. To model the 1D disordered cavity, we introduce random fluctuations of the refractive index. The resonator is divided into $100$ slices $100$~nm long each. The refractive index of each slice is set to $n_i = n_0 \left(1 + \sigma \xi_i \right)$, where $\xi_i$ is a random number that is uniformly distributed in the interval $\left[-1,1\right]$, and $\sigma$ is a free parameter used to tune the amount of disorder. For the results in Figs.~\ref{fig:activeModes}, B, D and F, as well as Figs.~\ref{fig:randCavModes} and \ref{fig:ActiveSimDisordered}, the value of $\sigma$ is $0.3$. Employing the Transfer Matrix Method, we calculated the transmission spectrum through the slab as well as the wavelengths and intensity distributions of resonant modes in the passive system, which are displayed in Fig.~\ref{fig:randCavModes}.

\subsection*{Time domain simulations of lasing dynamics}

We simulated the dynamics of the coupled electromagnetic field and semiconductor gain material on the basis of a full-wave time-domain model integrated into a finite-difference time-domain (FDTD) method through an auxiliary equation approach \cite{Boehringer2008Both}. The spatially dependent electric and magnetic fields are evolved in time by solving Maxwell's equations in their full form (i.e., beyond the slowly varying envelope approximation) on a discrete space and time grid according to the Yee FDTD scheme with a grid constant $\Delta x = 20$~nm and a time step $\Delta t = 0.0667$~fs. The interactions with the semiconductor gain medium are introduced through the auxiliary field
$\mathbf{D}\left(\mathbf{r},t\right) = \varepsilon_0 \varepsilon_b\left(\mathbf{r}\right) \mathbf{E}\left(\mathbf{r},t\right) + \mathbf{P}\left(\mathbf{r}, t\right)$,
where $\varepsilon_b$ is the static response and $\mathbf{P}\left(\mathbf{r},t\right)$ is the electronic contribution to the semiconductor polarization. 

The semiconductor gain medium is distributed over the entire cavity of $10~\mu$m length and refractive index $n_0 = 3.5$. Outside the cavity, the refractive index is $n = 1$. The semiconductor band structure is approximated by using the effective masses (in units of the electron mass $m_0$) listed in Table~\ref{tab:params}, and the envelope function approximation (EFA) is used to obtain renormalized quantum well values for the band gap and dipole matrix element starting from the corresponding bulk values. 

The time evolution of the polarization is obtained via a band-resolved density matrix approach to the electron dynamics in the semiconductor quantum well \cite{Boehringer2008Both,Buschlinger2015,Guazzotti2016}, which accounts for occupations as well as coherences between valence and conduction states. Due to the small momentum carried by the electromagnetic fields, only coherences between states with the same momentum $\mathbf{k}$ need to be considered. The resulting equations of motion for the microscopic polarizations, $p_\mathbf{r}\left(\mathbf{k},t\right)$, show a parametric dependence on the spatial position, $\mathbf{r}$, and couple to the electric field, $\mathbf{E}(\mathbf{r}, t)$, beyond the rotating wave approximation. The explicit inclusion of the semiconductor band structure obtained by modeling polarizations at different points of the reciprocal space provides a more accurate reproduction of the quantum well gain spectrum as compared to simpler models like four-level systems. An additional term is included at a macroscopic level to represent electrical pumping of the semiconductor at constant current~\cite{Hess1996,Boehringer2008Both}. We also apply a quasi-equilibrium approximation to the electron dynamics and locally assume a Fermi-Dirac distribution, $f_{N}\left(\mathbf{k}\right)$, parametrized by the density of carriers, $N\left(\mathbf{r},t\right)$, whose dynamics includes optical gain and absorption as well as electric pumping and non-radiative decay. 

\renewcommand{\thetable}{S\arabic{table}}

\begin{table}[tb]
\begin{center}
	\begin{tabular}{lcc}
		\hline
		\hline
		effective mass electrons 	& $m_e$		 & $0.063~m_0$\\
		effective mass holes 		& $m_h$      & $0.51~m_0$\\
		band gap 					& $E_\mathrm{gap}$	 & $1.4643$~eV\\
		dipole matrix element 		& $d$ & $0.5$~nm\\
		dephasing rate				& $\gamma$	& $12$~ps$^{-1}$ \\
		average refractive index & $n_0$ & $3.5$\\
		refractive index variation & $\sigma$ & $0.3$\\
		non-radiative decay & $\gamma_{nr}$ & $0.1$~ns$^{-1}$\\
		cavity length & $L$ & $10~\mu$m\\
		band structure discretization & $N_k$ & 31 \\
		\hline
		\hline
	\end{tabular}
\end{center}	
\caption{Values of the parameters used in the simulations of semiconductor lasers.}\label{tab:params}
\end{table}


\section*{Supplementary Text}

\subsection*{Dynamic filamentation in broad-area edge-emitting lasers}

\begin{figure*}[tb]
\begin{center}
\includegraphics[width = 13.0 cm]{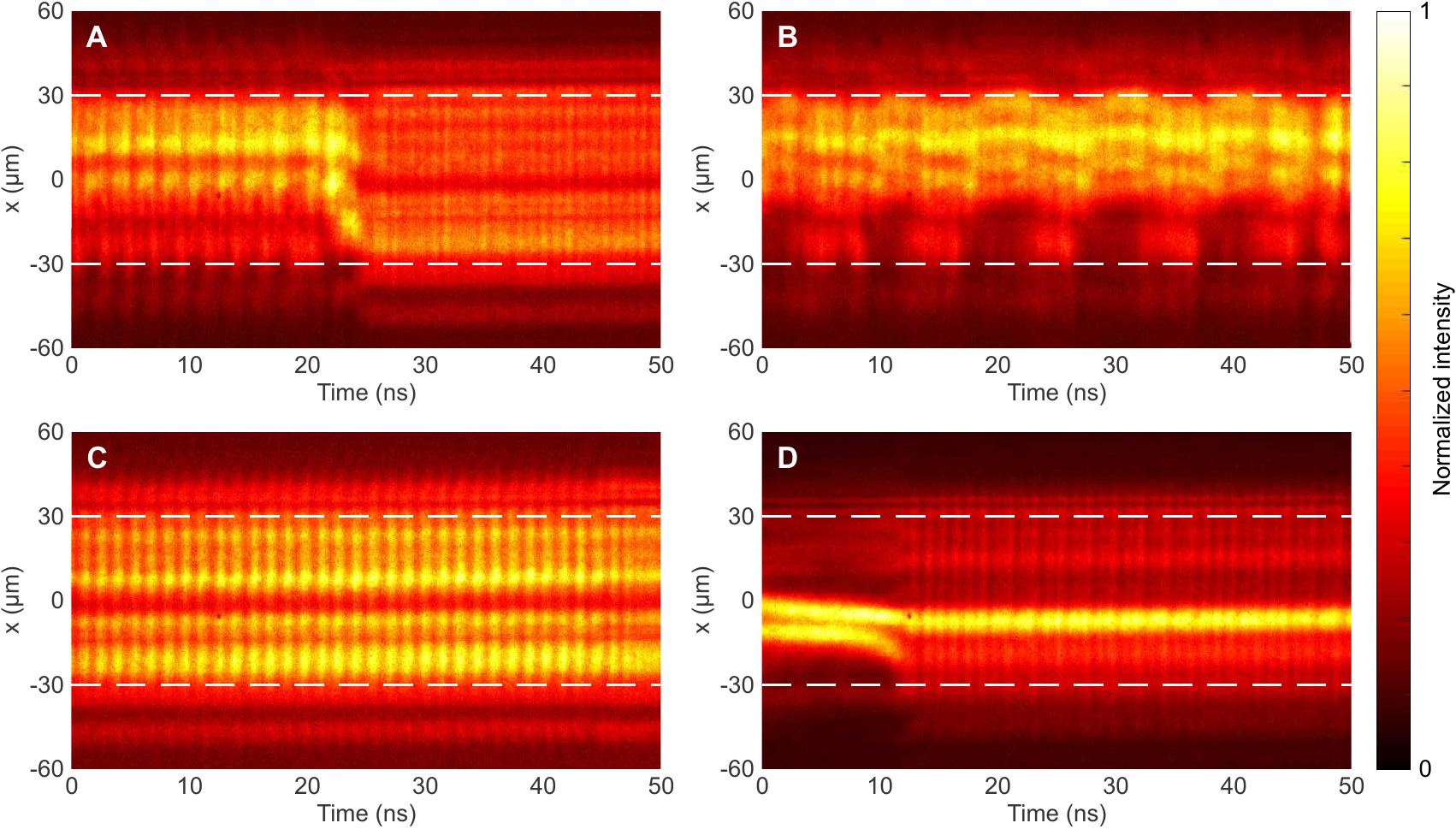}
\end{center}
\caption{Spatio-temporal traces of the emission intensity at an end facet of a $60~\mu$m-wide stripe laser over the course of a $2~\mu$s-long pulse. The dashed white lines denote the boundary of the top metal contact. The nonlinear dynamics leads to inhomogeneous spatial intensity profiles and temporal pulsations. The intensity distributions and pulsation frequencies can change suddenly during the pulse, highlighting the instability of the lasing dynamics.}
\label{fig:FPstreakExmpls}
\end{figure*}

Figure~\ref{fig:FPstreakExmpls} presents several examples of measured spatio-temporal traces of the emission intensity for the same $60~\mu$m-wide stripe laser shown in Fig.~\ref{fig:FPdynamics}. They were all measured over the course of a $2~\mu$s-long pulse. Different spatial intensity profiles and temporal pulsation patterns are observed. The lasing emission can either be distributed almost evenly over the end facet or be concentrated in certain locations. In addition, part of the emission stems from the regions outside of the top contact (marked by white dashed lines) since current can spread laterally in the GaAs cladding layer. Both the spatial profile and the temporal oscillation frequencies can suddenly change as seen in Figs.~\ref{fig:FPstreakExmpls}, A and D. These sudden changes highlight the instability of the nonlinear dynamics of the stripe laser. 

\begin{figure*}[tb]
\begin{center}
\includegraphics[width = 16.0 cm]{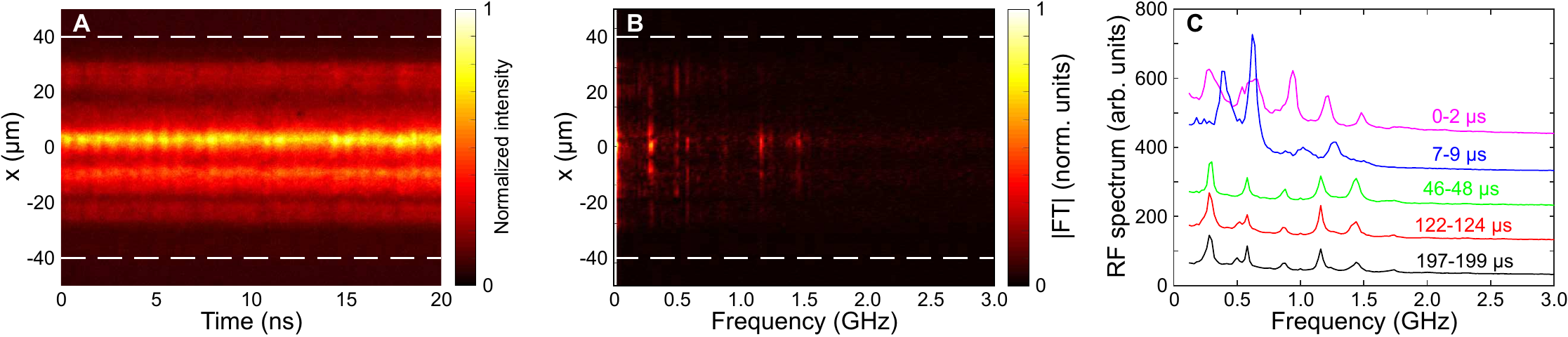}
\end{center}
\caption{Dynamics of a $80~\mu$m-wide and $1.02$~mm-long stripe laser pumped by $200~\mu$s-long pulses at $400$~mA current. (\textbf{A})~Spatio-temporal image of the emission intensity at $197~\mu$s after the start of the pump pulse and (\textbf{B})~its spatially resolved Fourier spectrum. The laser exhibits irregular pulsations with RF frequency components up to $1.5$~GHz. The dashed white lines indicate the boundaries of the top contact. (\textbf{C}) RF spectra of the emission intensity calculated from the spatio-temporal images, for five time periods during a $200~\mu$s-long pump pulse. Magenta: $0$--$2~\mu$s; blue: $7$--$9~\mu$s; green: $46$--$48~\mu$s; red: $122$--$124~\mu$s, black: $197$--$199~\mu$s. The curves are plotted with a vertical offset of $100$.}
\label{fig:FPlongPulse}
\end{figure*}

In such a broad-area edge-emitting laser, self-induced structure formation such as filamentation can be triggered by a modulational instability. As described in the main text, the filamentation is a direct consequence of a concert of microscopic processes. Due to the fast microscopic Coulomb-scattering processes and the link with the optical field dynamics, the relaxation and transport of interband polarization are faster than that of the (macroscopic) carrier density. This leads to a concentration of the field intensity in an index-guiding channel formed by local depletion of the carrier density. In the long and wide stripe broad-area cavity, a filament is longitudinally homogeneous, resulting in a wave-like reflection from the end facet. Due to spatially non-uniform field- and scattering-driven transport of carriers, the front of the filament becomes longitudinally and laterally more and more inhomogeneous. At the same time, the local carrier density outside the filament is not being depleted by stimulated emission and, consequently, fosters additional filaments. The filaments interact nonlinearly via the semiconductor gain medium, thereby destabilizing the lasing dynamics. The combination of a transverse modulational instability with the propagation of the filaments as well as the carrier transport then starts and sustains the filament migration. Thus, these spatio-temporal instabilities are intrinsic to the dynamics of broad-area semiconductor lasers. 

The temperature drift during the pulse also contributes to changes in the spatio-temporal dynamics. We further investigated the lasing dynamics with $200~\mu$s-long pump pulses in order to determine if steady-state lasing can be reached when the sample temperature stabilizes. Here, we present data for a $80~\mu$m-wide and $1.02$~mm-long stripe laser pumped with $400$~mA current, where the lasing threshold was at $I_{th} = 330$~mA. Spatio-temporal measurements were made for five $2~\mu$s-long intervals spread over the $200~\mu$s-long pulses, $0$--$2~\mu$s, $7$--$9~\mu$s, $46$--$48~\mu$s, $122$--$124~\mu$s, and $197$--$199~\mu$s. Figure~\ref{fig:FPlongPulse}A shows the spatio-temporal image of the emission intensity measured at $197~\mu$s after the start of the pulse. The image displays the typical unstable dynamics with several filaments and rapid pulsations. The pulsations are irregular, and the spatially resolved Fourier Transform (FT),
\begin{equation} \tilde{I}(x, f) = \left| \int dt \, I(x, t) \, e^{-2 \pi i f t} \, \right| \end{equation}
in Fig.~\ref{fig:FPlongPulse}B shows a broad spectrum extending up to $1.5$~GHz with dominant frequency components around $0.25$~GHz. The DC component of the spectrum is not shown. 

For each interval of $2~\mu$s length, we calculated the radio-frequency (RF) spectrum of the lasing emission intensity 
\begin{equation} S(f) = \left< \tilde{I}(x, f) \right>_{x, t, p} \end{equation}
by averaging over the spatial position $x$, the time $t$, and three consecutive pulses $p = 1,2,3$. In total, the average over $3 \times 40$ spatio-temporal images of $50$~ns length each was calculated. 

Figure~\ref{fig:FPlongPulse}C shows the RF spectra for the five time intervals. The RF spectrum changes significantly from the $0$--$2~\mu$s to the $7$--$9~\mu$s interval, and then again to the $46$--$48~\mu$s interval. These qualitative changes of the pulsation dynamics coincide with changes of the spatial intensity distributions (not shown). However, after the $46$--$48~\mu$s interval, the RF spectrum and the temporal dynamics as a whole no longer change significantly, i.e., the $122$--$124~\mu$s and $197$--$199~\mu$s intervals have nearly the same RF spectra as the $46$--$48~\mu$s interval. This indicates the sample temperature stabilizes, but the emission pulsations persist. These results exclude thermal effects as the fundamental cause for the dynamical instabilities of the stripe lasers. 

\subsection*{Emission profiles of D-cavity lasers}

\begin{figure}[tb]
\begin{center}
\includegraphics[width = 6.0 cm]{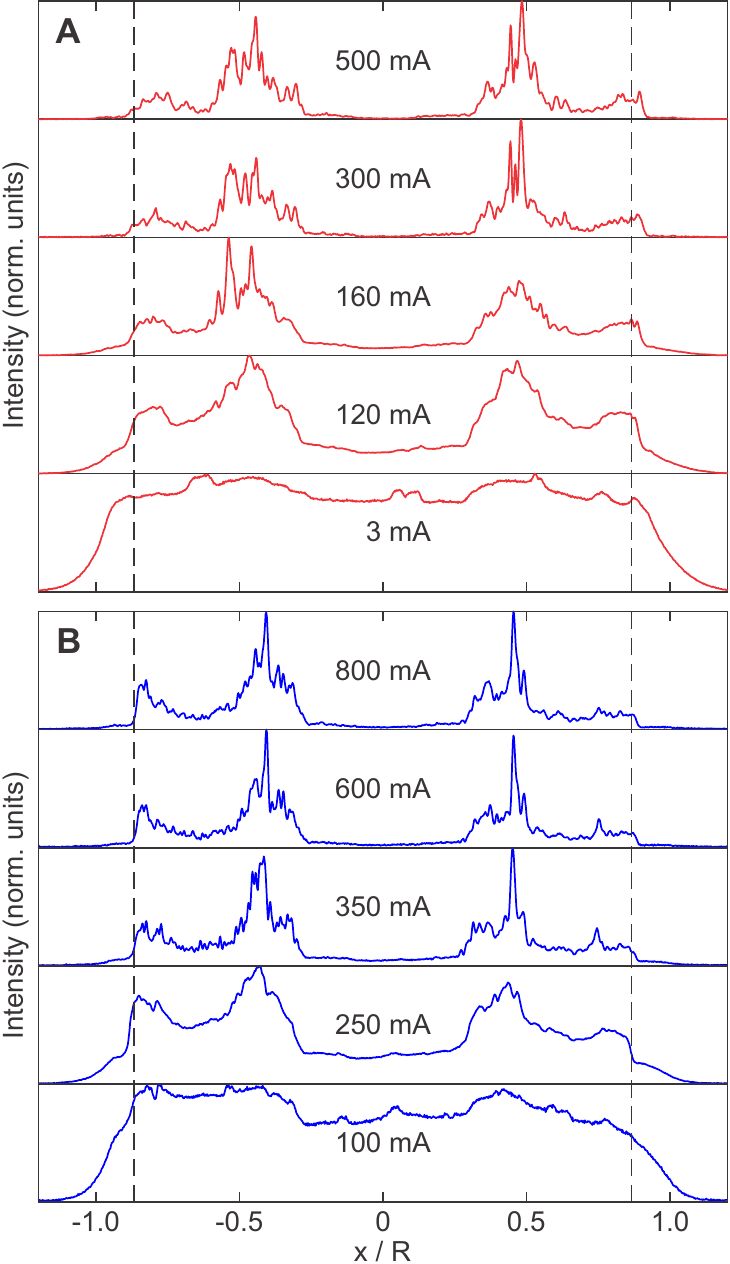}
\end{center}
\caption{Emission profiles on the straight segment of the boundary of two D-cavities fabricated by reactive ion etching. The intensity distributions were measured with a CCD camera for $2~\mu$s-long pulses. \textbf{(A)} Intensity distributions on the straight segment of a D-cavity with $100~\mu$m radius and pump currents $3$, $120$, $160$, $300$, and $500$~mA (from bottom to top), where the threshold is at $I_{th} = 150$~mA. \textbf{(B)} Intensity distributions on the straight segment of a D-cavity with $200~\mu$m radius and pump currents $100$, $250$, $350$, $600$, and $800$~mA (from bottom to top), where the threshold is at $I_{th} = 300$~mA. The vertical dashed black lines mark the corners of the D-cavities.}
\label{fig:DcavSWdistrExp}
\end{figure}

The emission intensity distributions on the straight segment of the boundary of the D-cavities (also called emission profiles) display an inhomogeneous structure as can be seen in the spatio-temporal traces in Figs.~\ref{fig:DcavDynamics} and \ref{fig:DcavDynamicsLong}. These cavities were fabricated by reactive ion etching and have high quality, e.g., the sidewalls are vertical and smooth as seen in the SEM images (Figs.~\ref{fig:SEM-Dcav}, A and B). To investigate the origin of the spatial inhomogeneity of the emission profiles, images of the intensity distributions integrated over $2~\mu$s-long pulses were taken for D-cavity lasers with radius $R = 100$ and $200~\mu$m, respectively. Figures~\ref{fig:DcavSWdistrExp}, A and B, show the intensity distributions for different pump currents of the two D-cavities presented in Figs.~\ref{fig:DcavDynamics}~and~\ref{fig:DcavDynamicsLong}, respectively. A $20\times$ microscope objective ($\mathrm{NA} = 0.40$) and a $10\times$ objective ($\mathrm{NA} = 0.25$) were used for the D-cavity with $R = 100~\mu$m and with $R = 200~\mu$m, respectively, in order to adjust the image size with respect to the camera chip size. 

Well below the lasing threshold, the intensity distributions are quite, though not completely homogeneous. It should be noted that the non-vanishing emission intensity observed beyond the two ends of the straight segment (marked by the black dashed lines in Fig.~\ref{fig:DcavSWdistrExp}) stems from the curved sidewalls, which are out of focus but still partially visible as can be seen in Fig.~\ref{fig:DcavDynamics}A. The spatially inhomogeneous intensity distribution develops with increasing pump current in Fig.~\ref{fig:DcavSWdistrExp}. Above the lasing threshold, the emission profile features a gap with comparatively little emission in the middle, surrounded by two regions of intense emission on both sides, and two further regions of weaker emission near the corners. Narrow peaks appear and disappear in the regions of strong emission as the pump current increases, but the overall structure stays the same. 

Therefore, the emission profiles of the D-cavity lasers exhibit two distinct length scales. The coarse scale is the size of the regions of strong or weak emission, and is of the order of tens of micrometers. The fine scale is the width of the narrow peaks inside the bright regions and is of the order of a few micrometers. 

The coarse-scale structure of the emission pattern is identical for D-cavity lasers of different size, and its length scale is proportional to the cavity size as shown in Fig.~\ref{fig:DcavSWdistrExp}. Hence, this structure is not formed by the nonlinear interaction of the optical field with the gain medium, which would determine the length scale of the filaments irrespective of the cavity size. Furthermore, filamentation dynamics typically leads to rapid pulsations on the nanosecond time scale, which are absent for the D-cavity lasers. 

Instead, the intensity distributions shown in Fig.~\ref{fig:DcavSWdistrExp} result from the resonant modes of the D-cavities. They actually correspond to an incoherent sum of emission profiles of high-$Q$ modes, as will be shown below. These modes experience stronger amplification due to their longer lifetime, thus grow faster with increasing pump current and contribute more and more to the emission intensity, leading to the transition of the emission profiles in Fig.~\ref{fig:DcavSWdistrExp}. Above the threshold, the coarse-structure of the emission profiles no longer changes, and it is determined by the high-$Q$ modes which become the lasing modes. 

On top of the coarse structure discussed above, there is a fine structure of peaks only a few micrometers wide. These narrow peaks change as the pump is increased (see Fig.~\ref{fig:DcavSWdistrExp}) as well as during the emission pulses (see Figs.~\ref{fig:DcavDynamics} and \ref{fig:DcavDynamicsLong}). They stem from the intensity distributions of individual lasing modes that turn on and off during the pulse due to temperature changes or as the pump current increases. Since the mode structures consist of speckle grains of the order of the wavelength, the size of these peaks is of the same order and cannot be resolved by our imaging setup. In fact, the width of the peaks in the measured intensity profiles is given by the resolution limit of the microscope objectives. 

\begin{figure*}[tb]
\begin{center}
\includegraphics[width = 12 cm]{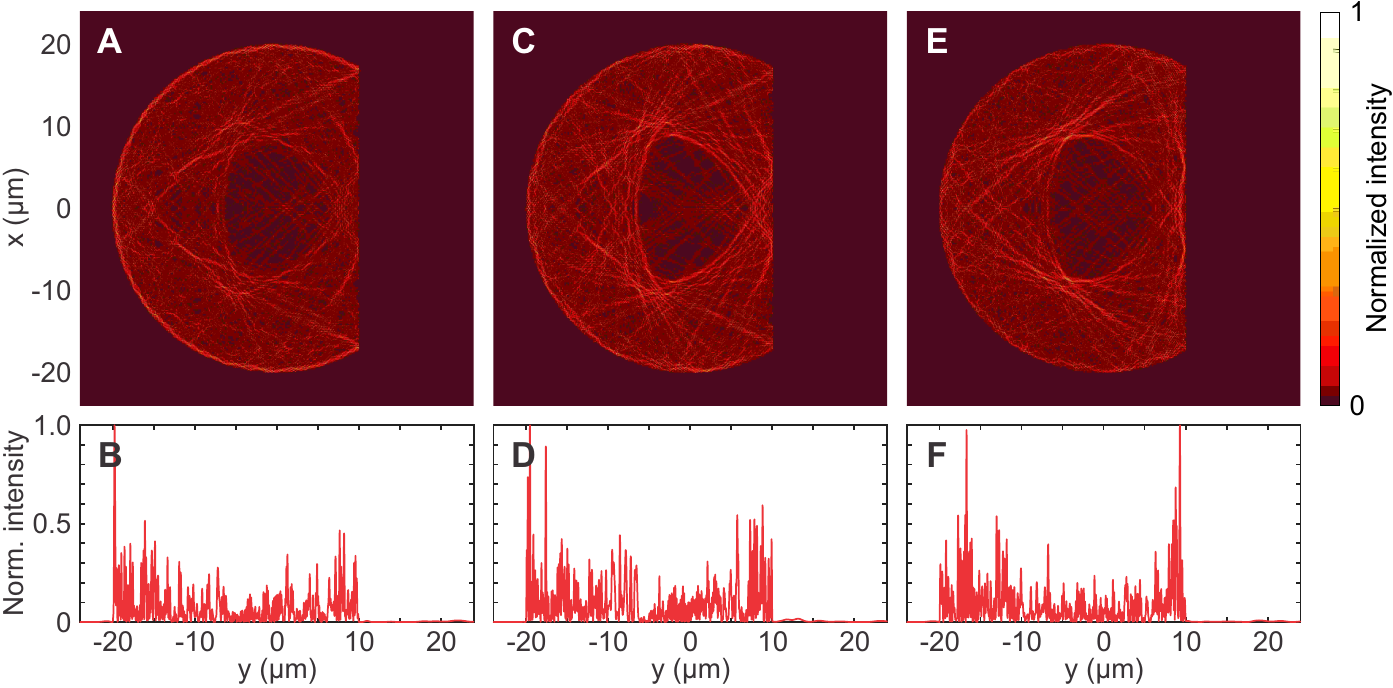}
\end{center}
\caption{\textbf{(A, C, E)} Intensity distributions and \textbf{(B, D ,F)} cuts at $x = 0$ of three modes of a D-cavity with radius $R = 20~\mu$m and refractive index $n = 3.37$. \textbf{(A, B)}: a mode with $\lambda = 800.4$~nm and $Q = 3531$, \textbf{(C, D)}: a mode with $\lambda = 800.0$~nm and $Q = 3699$, \textbf{(E, F)}: a mode with $\lambda = 799.6$~nm and $Q = 3170$.}
\label{fig:DcavModesCuts}
\end{figure*}

In order to compare the measured intensity distributions to the passive cavity modes, we calculated the modes of a D-cavity with $R = 20~\mu$m radius (see methods). Three examples of spatial intensity distributions for modes with quality factors $Q \geq 3000$, which are among the most long-lived modes of the cavity, are shown in Fig.~\ref{fig:DcavModesCuts}. While the details of their intensity distributions differ, they all show similar features in their coarse structure like a region of smaller than average intensity in the middle of the cavity. This can also be observed in the cuts along $x = 0$ shown in Figs.~\ref{fig:DcavModesCuts}, B, D, and F. 

\begin{figure*}[tb]
\begin{center}
\includegraphics[width = 12 cm]{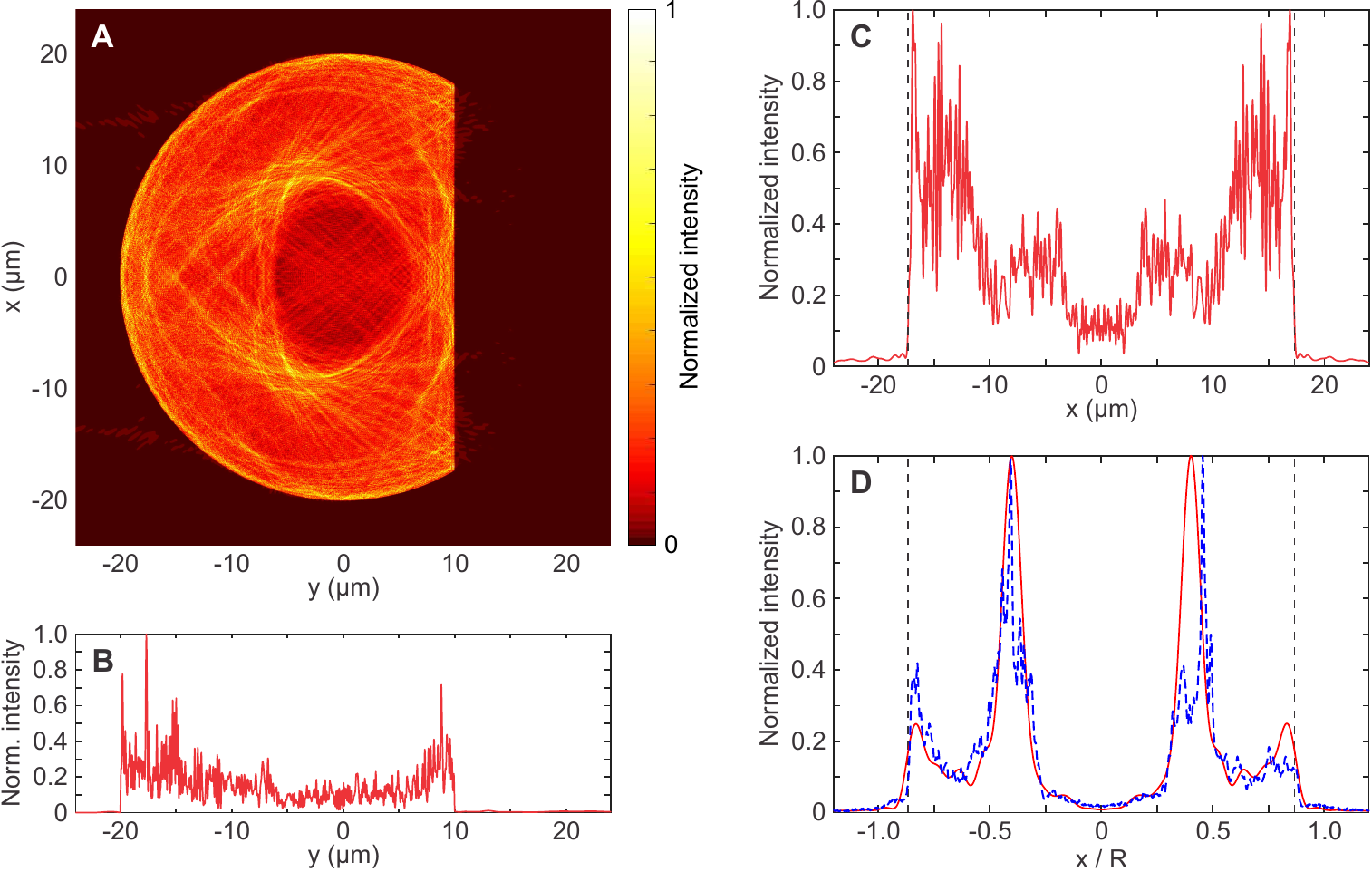}
\end{center}
\caption{\textbf{(A)} Sum of the intensity distributions of eleven modes with quality factors $Q = 3000$--$4000$ of a D-cavity with radius $R = 20~\mu$m and refractive index $n = 3.37$ and \textbf{(B)} cut at $x = 0$ through it. \textbf{(C)} The sum emission profile at the straight segment of the boundary ($y = R/2$) and \textbf{(D)} the sum emission profile (red solid line) after convolution with the point spread function of the objective ($\mathrm{NA} = 0.40$), which agrees well to an experimentally measured emission profile (blue dashed line) for a D-cavity with $R = 200~\mu$m radius pumped with a current of $800$~mA (cf.\ Fig.~\ref{fig:DcavSWdistrExp}B). The spatial coordinate $x$ is normalized by the radius $R$. The vertical dashed lines in (C) and (D) indicate the corners of the D-cavity.}
\label{fig:DcavModeSum}
\end{figure*}

With optical gain added to the cavity, the high-$Q$ modes within the gain spectrum will lase first due to their lower thresholds. Assuming all modes lase independently without phase coherence, the total emission profile is an incoherent sum of the emission profiles of individual modes. Numerically we sum the intensity distributions for $11$ modes with the highest quality factors ($3000 \leq Q \leq 4000$) in the relevant wavelength range for GaAs quantum well emission. The sum intensity distribution in Fig.~\ref{fig:DcavModeSum}A displays similar features as the individual modes in Fig.~\ref{fig:DcavModesCuts}. 

For a direct comparison with the experimentally measured intensity distributions in Fig.~\ref{fig:DcavSWdistrExp}, the sum intensity distribution at the straight segment of the boundary is plotted in Fig.~\ref{fig:DcavModeSum}C. The sum emission profile is indeed inhomogeneous, but not identical to the measured intensity distributions in Fig.~\ref{fig:DcavSWdistrExp}. The difference is caused by the finite spatial resolution of the imaging optics. To account for the numerical aperture of $\mathrm{NA} \leq 0.40$ of the objective used in the experiment, the field distributions of the individual modes were Fourier-transformed into momentum-space, where a rectangular filter of full width $\mathrm{NA} \, 2 \pi / \lambda$ was applied, before transforming them back into real space. After adding their emission profiles incoherently, the total emission profile, plotted by the red line in Fig.~\ref{fig:DcavModeSum}D (and as red line in Fig.~\ref{fig:DcavDynamics}F), agrees well with the experimental data plotted by the blue line (identical to the top-most curve in Fig.~\ref{fig:DcavSWdistrExp}B). This agreement confirms that the measured emission intensity distributions are determined by the passive cavity modes in the D-cavity, instead of the nonlinear interaction of the field with the gain medium as in the case of the stripe lasers. 

It should be noted that the calculated intensity distribution in Fig.~\ref{fig:DcavModeSum}D does not reproduce the fine features of the measured one. First, the cavities used in the experiments are five to ten times larger than what is considered in the simulations, and hence the width of a resolution limited peak, which is of the order of $1~\mu$m, is much larger in terms of the radius in the case of the simulations. Second, the fine features depend on the exact combination of cavity modes that lase, which in turn depends on small boundary variations and the sample temperature. Therefore the fine structures vary over the course of a pump pulse and also from cavity to cavity. 

\subsection*{Spatially-resolved lasing spectra}

\begin{figure}[tb]
	\begin{center}
		\includegraphics[width = 8.4 cm]{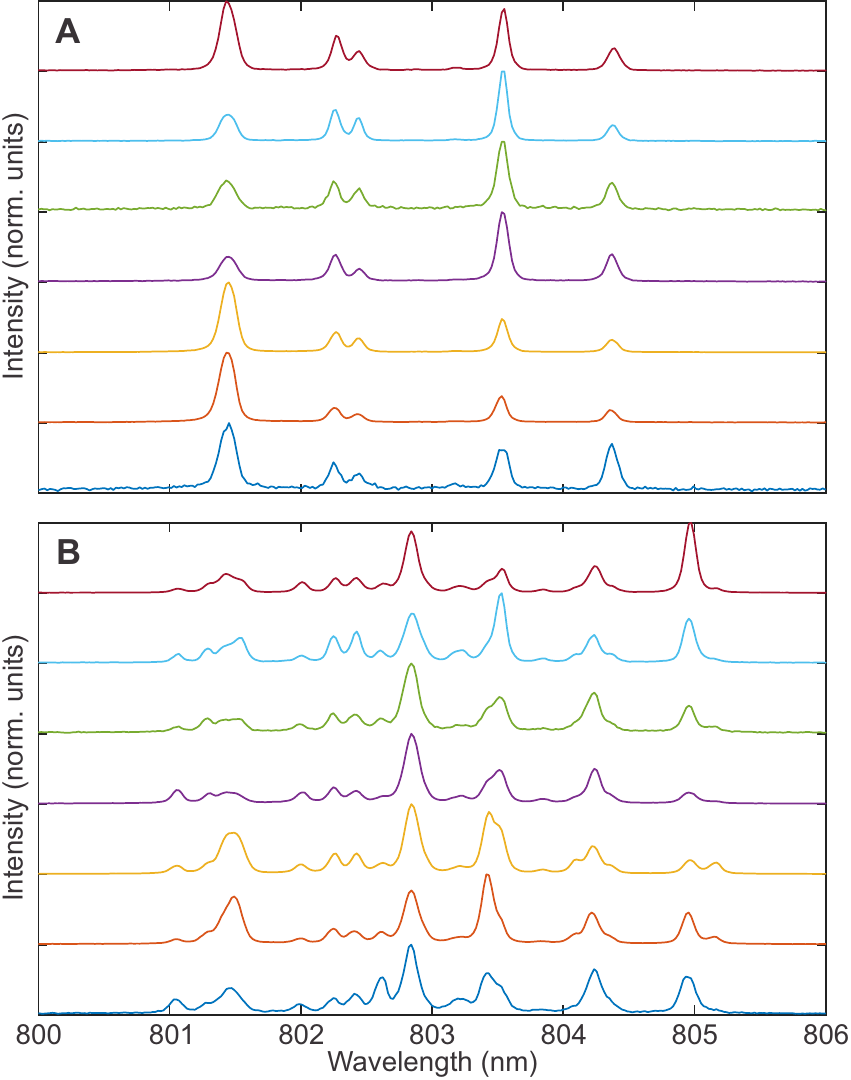}
	\end{center}
	\caption{Spatially-resolved emission spectra of a D-cavity laser fabricated by reactive ion etching with $R = 100~\mu$m radius. Each spectrum corresponds to the emission transmitted by one fiber of the fiber bundle. The spectra are normalized to unit amplitude. Exemplary spectra for \textbf{(A)} $240$~mA and \textbf{(B)} $460$~mA pump current are shown, where the threshold current is $I_{th} = 150$~mA.}
	\label{fig:DcavSpatioSpectral}
\end{figure}

To investigate correlations between the emission spectra from different parts of the cavity boundary, we measured the lasing spectra of a D-cavity with spatial resolution (see methods). The cavity is fabricated by reactive ion etching and has a radius of $R = 100~\mu$m. Figures~\ref{fig:DcavSpatioSpectral}, A and B, show the spectra of lasing emission from seven different sections of the straight segment of the boundary of the D-cavity at two pump currents of $240$ and $460$~mA. Each spectrum consists of multiple peaks, where the maximum of each spectrum is normalized to $1$. All the peaks are present in the emission from all seven sections, though their relative heights vary from one location to another. This observation is consistent with the numerical simulations showing that the high-$Q$ modes of the D-cavity are spatially distributed throughout the cavity (Fig.~\ref{fig:DcavModesCuts}). Although the emission from the center of the straight segment is relatively weak, it still contains all the peaks. Therefore, there is no significant correlation between the shape of the spectrum and the spatial location of emission on the cavity boundary. 

It should be mentioned that wave-chaotic cavities can also support modes localized on unstable periodic orbits, so-called scars \cite{Cao2015}, which may have different effects on the lasing dynamics \cite{Sunada2005}. However, in the D-cavity, the scar modes have rather low quality factors and thus do not contribute to lasing. 

\subsection*{Effect of boundary roughness on D-cavity lasers}

\begin{figure*}[tb]
	\begin{center}
		\includegraphics[width = 11 cm]{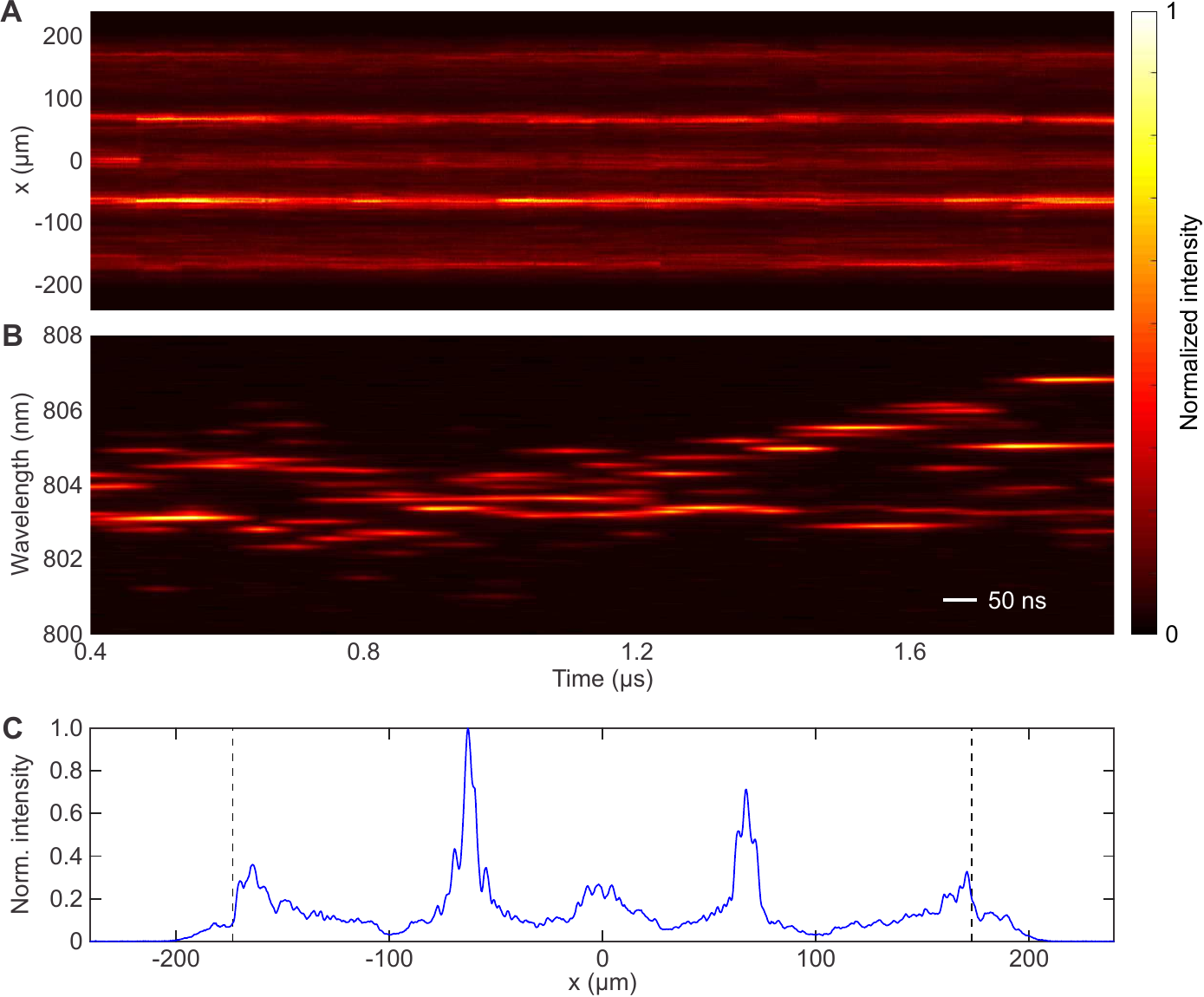}
	\end{center}
	\caption{Lasing dynamics of a D-cavity fabricated by wet chemical etching with radius $R = 200~\mu$m. \textbf{(A)}~Spatio-temporal image of emission intensity for $1200$~mA pump current during the interval $1.4$--$1.9~\mu$s of a $2~\mu$s-long pulse, where the threshold current is $I_{th} = 470$~mA. \textbf{(B)}~Spectrochronogram measured with the same pump conditions as (A) with $50$~ns resolution. \textbf{(C)}~Emission intensity profile on the straight segment of the boundary of the D-cavity for $1200$~mA pump current integrated over a $2~\mu$s-long pulse.}
	\label{fig:DcavDynamicsWet}
\end{figure*}

In order to study the effects of fabrication imperfections on the lasing dynamics in D-cavities, we tested the samples fabricated by wet chemical etching (see methods) and compared them to those fabricated by reactive ion etching. The wet-etched cavities have non-vertical and relatively rough sidewalls (Fig.~\ref{fig:SEM-Dcav}D), and their lasing threshold currents are about $1.5$--$3$ times higher than those of the dry-etched D-cavities with the same size. 

Nevertheless, the spatio- and spectro-temporal dynamics of the wet-etched D-cavity lasers are qualitatively identical to those of the dry-etched ones. Figure~\ref{fig:DcavDynamicsWet} shows the spatio-temporal image and the spectrochronogram of a wet-etched D-cavity with $R = 200~\mu$m. The spectral and spatial correlation times are $\tau_\mathrm{corr}^{(\lambda)} = 72$~ns and $\tau_\mathrm{corr}^{(x)} = 50$~ns, respectively. These values are only a bit shorter than those of the dry-etched D-cavity with the same radius presented in Fig.~\ref{fig:DcavDynamicsLong}. The slight difference is attributed to stronger heating of the wet-etched sample since a larger pump current was used due to its higher lasing threshold. 

Figure \ref{fig:DcavDynamicsWet}C shows the spatial distribution of emission intensity on the straight segment of the wet-etched D-cavity. The emission profile is also inhomogeneous, but different from that of the dry-etched cavities in Fig.~\ref{fig:DcavSWdistrExp}. This difference is caused by the changes in the spatial structures of the passive cavity modes due to the boundary roughness, which is more significant than for the dry-etched cavities.  

\begin{figure*}[tb]
	\begin{center}
		\includegraphics[width = 14 cm]{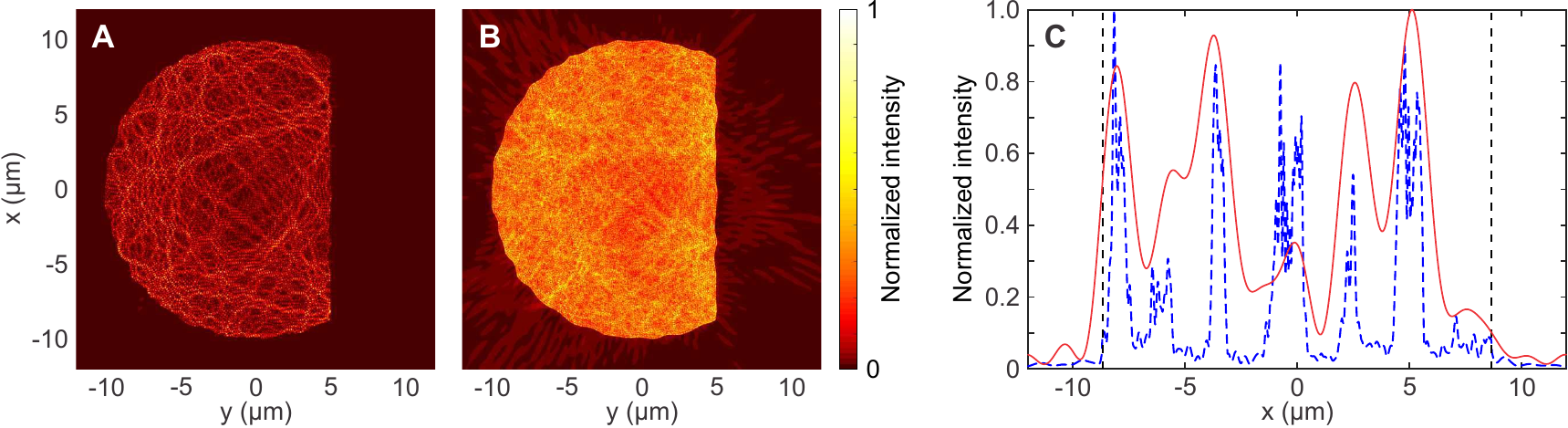}
	\end{center}
	\caption{D-cavity with boundary roughness ($R = 10~\mu$m, $n = 3.37$). \textbf{(A)} Intensity distribution of a typical high-$Q$ mode with $\lambda = 799.32$~nm and $Q = 1540$. \textbf{(B)} Sum of the intensity distributions of fifteen modes with quality factors $Q = 1100$--$1600$. \textbf{(C)} Sum of the intensity distributions at the straight segment of the boundary ($y = R/2$) without (blue dashed line) and with (red solid line) convolution with the point spread function of the objective ($\mathrm{NA} = 0.40$). The vertical dashed lines indicate the corners of the D-cavity.}
	\label{fig:DcavRoughModes}
\end{figure*}

We simulated the boundary roughness by adding perturbations to the cavity boundary (see methods). A typical example of a high-$Q$ mode of a D-cavity with boundary roughness ($R = 10~\mu$m, $n = 3.37$) is shown in Fig.~\ref{fig:DcavRoughModes}A. The speckle-like fine-scale structure is very similar to that of the modes of a D-cavity without boundary roughness (see Fig.~\ref{fig:DcavModesCuts}). On a larger scale, however, the intensity distribution is more homogeneous across the cavity, in particular the region of lower intensity in the middle of the D-cavity with smooth boundary is no longer visible. The sum of the intensity distributions of fifteen high-$Q$ modes is shown in Fig.~\ref{fig:DcavRoughModes}B. Like the individual mode in Fig.~\ref{fig:DcavRoughModes}A, there is no discernible large-scale structure, evidencing that the boundary roughness can indeed qualitatively change the intensity distributions of the lasing modes. This is also confirmed by the sum intensity distribution on the straight segment in Fig.~\ref{fig:DcavRoughModes}C, which is still inhomogeneous, but qualitatively different from that of a smooth D-cavity (cf.\ Figs.~\ref{fig:DcavModeSum}, C and D). After convolution with the point spread function of the objective ($\mathrm{NA} = 0.40$) used in the experiment, the emission profile agrees qualitatively with the measured intensity distribution of the wet-etched D-cavity in Fig.~\ref{fig:DcavDynamicsWet}C, featuring for example a central region of relatively strong emission. 

In spite of the differences in the spatial emission profiles, the wet-etched D-cavities feature stable lasing dynamics like the dry-etched ones. Therefore, the complex wave interference in the wave-chaotic cavities persists in the presence of boundary perturbations and fabrication imperfections, and remains effective in suppressing filamentation and lasing instabilities. 

\subsection*{One-dimensional semiconductor lasers}

\begin{figure}[tb]
\begin{center}
\includegraphics[width = 8.4 cm]{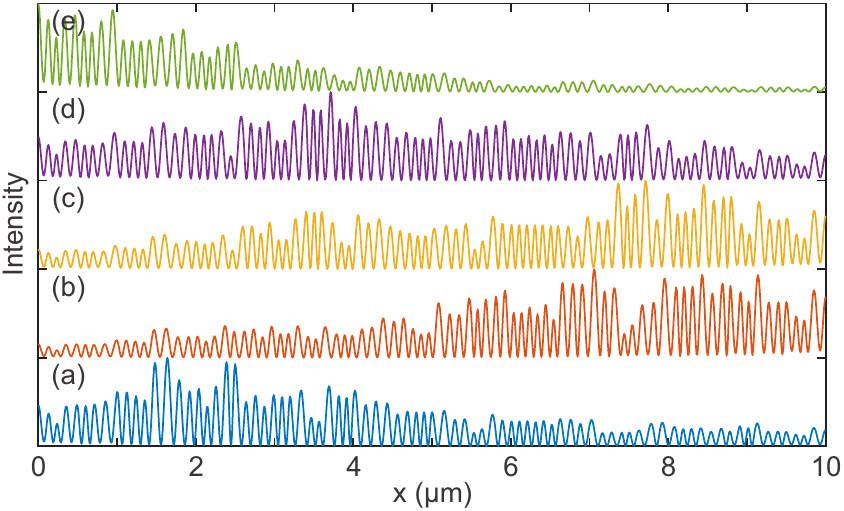}
\end{center}
\caption{Passive mode intensity distributions of the 1D disordered cavity. The cavity is $10~\mu$m long and the same used in the active simulations. The spatially-varying refractive index profile is shown in Fig.~\ref{fig:activeModes}B. The intensity distributions of five modes in the spectral region between $800$~nm and $850$~nm are shown, with the corresponding wavelengths and quality factors (from bottom to top) (a) $\lambda = 843.0$~nm and $Q = 274$, (b) $\lambda = 833.8$~nm $Q = 257$, (c) $\lambda = 821.6$~nm and $Q = 273$, (d) $\lambda = 814.0$~nm and $Q = 302$, and (e) $\lambda = 801.0$~nm and $Q = 124$.}
\label{fig:randCavModes}
\end{figure}

Figure~\ref{fig:randCavModes}A shows the intensity distributions of resonant modes in the disordered passive cavity (see Fig.~\ref{fig:activeModes}B) within the semiconductor quantum well gain spectrum. The intensity distributions have an irregular structure in contrast to those of the cavity with homogeneous refractive index profile (see Fig.~\ref{fig:activeModes}A). The modes are extended throughout the whole cavity with oscillations on the wavelength scale, however, their envelopes feature irregular fluctuations on much longer scales. In this respect, they are very similar to the modes of the wave-chaotic D-cavity in Fig.~\ref{fig:DcavModesCuts}. This becomes particularly apparent when comparing to the line cuts through the intensity distributions of the D-cavity shown in Figs.~\ref{fig:DcavModesCuts}, B, D, and F. Hence, the 1D disordered cavity is analogous to the D-cavity regarding the pseudo-random, speckle-like structure of the passive cavity modes. 

\begin{figure*}[p]
\begin{center}
\includegraphics[width = 11 cm]{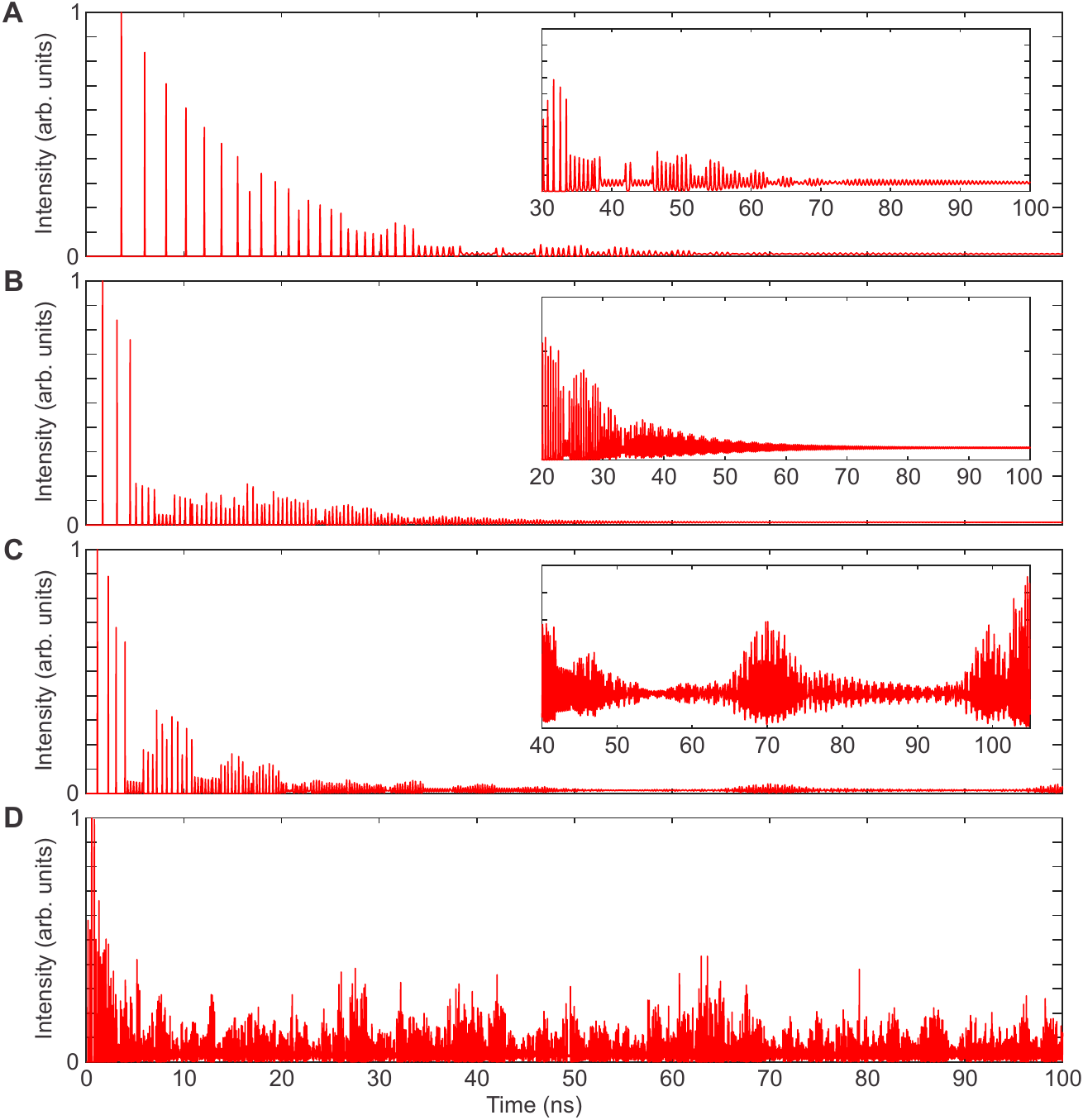}
\end{center}
\caption{Simulation of lasing dynamics in a $10~\mu$m-long 1D cavity with homogeneous refractive index profile for pump current densities \textbf{(A)}~$J = 110$~A cm$^{-2}$, \textbf{(B)}~$J = 130$~A~cm$^{-2}$, \textbf{(C)}~$J = 150$~A~cm$^{-2}$, and \textbf{(D)}~$J = 500$~A~cm$^{-2}$. The lasing threshold is $J_{th} \simeq 104$~A~cm$^{-2}$. The initial one or two peaks often go well above the intensity scale of the plot and are hence not fully displayed.}
\label{fig:ActiveSimHomogeneous}
\end{figure*}

We compare the lasing dynamics of one-dimensional lasers with homogeneous and disordered refractive index profile at various pumping levels. The cavities are the same as those presented in Fig.~\ref{fig:activeModes}. Figure~\ref{fig:ActiveSimHomogeneous} shows the total emission intensity of the homogeneous cavity for four different pump currents. For pump currents of $110$~A~cm$^{-2}$ and $130$~A~cm$^{-2}$ (Figs.~\ref{fig:ActiveSimHomogeneous}, A and B), just above the lasing threshold $J_{th} \simeq 104$~A~cm$^{-2}$, the lasing dynamics in the homogeneous cavity slowly stabilizes after a long transient (about $70$~ns). When increasing the pump current to $150$~A~cm$^{-2}$, the dynamics no longer stabilizes at all as shown in Fig.~\ref{fig:ActiveSimHomogeneous}C. The timescale of the pulsations decreases further without reaching stability when the pump current increases. This can be seen in the example for $500$~A~cm$^{-2}$ in Fig.~\ref{fig:ActiveSimHomogeneous}D, which shows a longer part of the time trace presented in Fig.~\ref{fig:activeModes}E. Therefore, the homogeneous cavity supports stable lasing dynamics only for a small range of pump currents close to the lasing threshold. 

\begin{figure*}[p]
\begin{center}
\includegraphics[width = 11 cm]{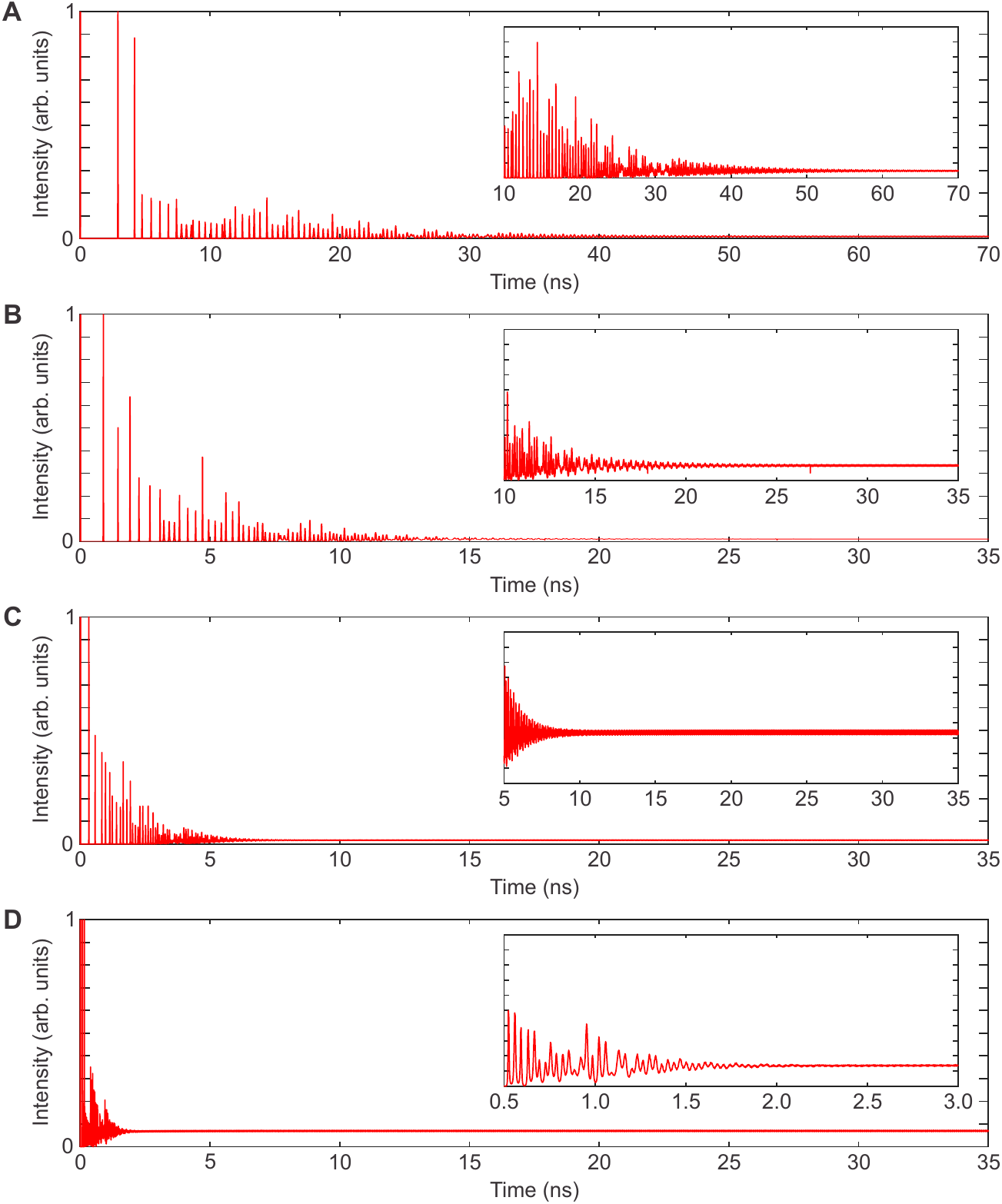}
\end{center}
\caption{Simulation of lasing dynamics in a $10~\mu$m-long one-dimensional disordered cavity (spatially-varying refractive index profile shown in Fig.~\ref{fig:activeModes}B) for pump current densities \textbf{(A)}~$J = 110$~A cm$^{-2}$, \textbf{(B)}~$J = 200$~A cm$^{-2}$, \textbf{(C)}~$J = 500$~A cm$^{-2}$, and \textbf{(D)}~$J = 2000$~A cm$^{-2}$. The threshold is at $J_{th} \simeq 96$~A~cm$^{-2}$. The first one or two peaks usually go well above the shown intensity scale and are hence not fully displayed.}
\label{fig:ActiveSimDisordered}
\end{figure*}

This is in sharp contrast to the disordered cavity, which reaches steady-state lasing after a transient. It should be noted that its threshold of $J_{th} \simeq 96$~A~cm$^{-2}$ is almost identical to that of the homogeneous cavity. Several examples of the lasing dynamics of the disordered cavity for different pump currents are shown in Fig.~\ref{fig:ActiveSimDisordered}, ranging from very close to the lasing threshold such as $J = 110$~A cm$^{-2}$ and  $J = 200$~A cm$^{-2}$ in Figs.~\ref{fig:ActiveSimDisordered}, A and B, to about five times of the threshold in Fig.~\ref{fig:ActiveSimDisordered}C ($J = 500$~A cm$^{-2}$, same data as in Fig.~\ref{fig:activeModes}F) and even twenty times of the threshold ($J = 2000$~A cm$^{-2}$) in Fig.~\ref{fig:ActiveSimDisordered}D. As the pump current is increased, the duration of the transient dynamics becomes shorter, and the stable emission sets in earlier. These results demonstrate the general validity and robustness of our scheme to achieve stable multimode operation of semiconductor lasers even for very high pump powers. 

It is important to mention that coherent instabilities have been reported in the numerical studies of 1D random lasers in the bad cavity limit \cite{Andreasen2011}. Our 1D disordered systems, however, are within the good cavity limit. Comparison of these two works indicates that if the cavity is too leaky, the multi-wave interference effect would not be sufficient to suppress lasing instabilities. However, most lasers employed in practical applications are in the good cavity limit, and our scheme of suppressing lasing instabilities is hence applicable. 

Finally we point out the differences between our 1D disordered cavity laser and the fiber laser with distributed feedback provided by Rayleigh scattering from the inhomogeneities in glass in Ref.~\cite{Turitsyn2010}. In the latter, fluctuations of the emission intensity appeared just above the lasing threshold, but disappeared at higher pumping level. The instabilities were caused by the combined effect of stimulated Brillouin scattering (SBS) and Rayleigh scattering. With increasing power, the SBS was suppressed since nonlinear interactions (multiple four-wave mixing processes) broadened the spectrum and reduced the power spectral density. However, SBS is negligible in our semiconductor lasers due to their short cavity length and relatively low quality factor. In addition, for the fiber random laser, light amplification is dominant over scattering, since the scattering length is much longer than the gain length. In our disordered cavity, in contrast, the refractive index fluctuates on a length scale of $100$~nm, which is much shorter than the gain length ($8.5~\mu$m at the lasing threshold). Light scattering occurs so frequently on a sub-wavelength scale that it effectively perturbs the coherent nonlinear processes that would lead to instabilities. 

\end{document}